\documentclass[
  journal=pasa,
  manuscript=research-paper, %% or "review"
  year=2026,
  volume=YY,
]{cup-journal}

\usepackage{amsmath}
\usepackage{amssymb,microtype,siunitx,booktabs}
\usepackage{caption}
\usepackage{subcaption}
\usepackage{orcidlink}
\usepackage{rotating}
\newcommand{\logniiha}{[N\textsc{ii}]/H$\alpha$}
\newcommand{\logoiiihb}{[O\textsc{iii}]/H$\beta$}
\newcommand{\logoiihb}{[O\textsc{ii}]/H$\beta$}
\newcommand{\logmstar}{$\rm M_*$}
\newcommand{\radioflux}{$S_{1.4\,{\rm GHz}}$}
\usepackage{multirow}
\usepackage{hyperref}
\hypersetup{colorlinks=true,
            linkcolor=blue,
            urlcolor=blue,
            linktoc=all,
            citecolor=blue}
\title{EMU/GAMA: A statistical perspective on active galactic nuclei diagnostics}

\author{J. Prathap\textsuperscript{\orcidlink{0009-0004-0251-2672}}}
\affiliation{School of Mathematical and Physical Sciences, Macquarie University, Sydney, NSW 2109, Australia}
\alsoaffiliation{Macquarie University Astrophysics and Space Technologies Research Centre, Sydney, NSW 2109, Australia}
\alsoaffiliation{Australia Telescope National Facility, CSIRO, Space and Astronomy, PO Box 1130, Bentley, WA 6102, Australia}
\email[J. Prathap]{jahangprathap12@gmail.com}

\author{A. M. Hopkins\textsuperscript{\orcidlink{0000-0002-6097-2747}}}
\affiliation{School of Mathematical and Physical Sciences, Macquarie University, Sydney, NSW 2109, Australia}
\alsoaffiliation{Macquarie University Astrophysics and Space Technologies Research Centre, Sydney, NSW 2109, Australia}

\author{R. Carvajal\textsuperscript{\orcidlink{0000-0002-0545-1113}}}
\affiliation{Instituto de Astrofísica e Ciências do Espaço, Universidade de Lisboa, OAL, Tapada da Ajuda, PT1349-018 Lisbon, Portugal}
\alsoaffiliation{Departamento de F\'isica, Faculdade de Ci\^encias, Universidade de Lisboa, Edif\'ecio C8, Campo Grande, PT1749-016 Lisbon}

\author{M. Cowley\textsuperscript{\orcidlink{0000-0002-4653-8637}}}
\affiliation{School of Chemistry and Physics, Queensland University of Technology, Brisbane, QLD, Australia}
\alsoaffiliation{Centre for Astrophysics, University of Southern Queensland, West Street, Toowoomba, QLD 4350, Australia}

\author{S. M. Croom\textsuperscript{\orcidlink{0000-0003-2880-9197}}}
\affiliation{Sydney Institute for Astronomy (SIfA), School of Physics, A28, The University of Sydney, NSW 2006, Australia}

\author{D. Farrah\textsuperscript{\orcidlink{0000-0003-1748-2010}}}
\affiliation{Department of Physics and Astronomy, University of Hawai’i at Mānoa, 2505 Correa Rd., Honolulu, HI 96822, USA}
\alsoaffiliation{Institute for Astronomy, University of Hawai’i, 2680 Woodlawn Dr., Honolulu, HI 96822, USA}

\author{I. Prandoni\textsuperscript{\orcidlink{0000-0001-9680-7092}}}
\affiliation{INAF – Istituto di Radioastronomia, Via P. Gobetti 101, 40129 Bologna, Italy}

\author{S. S. Shabala\textsuperscript{\orcidlink{0000-0001-5064-0493}}}
\affiliation{School of Natural Sciences, University of Tasmania, Private Bag 37, Hobart, Tasmania 7001, Australia}
 
\author{J. Th. van Loon\textsuperscript{\orcidlink{0000-0002-1272-3017}}}
\affiliation{Lennard-Jones Laboratories, Keele University, ST5 5BG, UK}

\author{C. Pappalardo\textsuperscript{\orcidlink{0000-0003-2606-6019}}}
\affiliation{Instituto de Astrofísica e Ciências do Espaço, Universidade de Lisboa, OAL, Tapada da Ajuda, PT1349-018 Lisbon, Portugal}
\alsoaffiliation{Departamento de F\'isica, Faculdade de Ci\^encias, Universidade de Lisboa, Edif\'ecio C8, Campo Grande, PT1749-016 Lisbon}

\author{K. A. Pimbblet\textsuperscript{\orcidlink{0000-0002-3963-3919}}}
\affiliation{E. A. Milne Centre for Astrophysics, University of Hull, Kingston-upon-Hull, UK}
\alsoaffiliation{Centre of Excellence for Data Science, AI, and Modelling (DAIM), University of Hull, Kingston-upon-Hull, UK}

\author{U. T. Ahmed\textsuperscript{\orcidlink{0000-0002-0309-1599}}}
\affiliation{Australian Astronomical Optics, Macquarie University, 7-9 Wally’s Walk, Sydney, NSW 2109, Australia}
\alsoaffiliation{Macquarie University Astrophysics and Space Technologies Research Centre, Sydney, NSW 2109, Australia}
\alsoaffiliation{Centre for Astrophysics, University of Southern Queensland, 37 Sinnathamby Boulevard, Springfield Central, QLD 4300, Australia}

\author{M. Bilicki\textsuperscript{\orcidlink{0000-0002-3910-5809}}}
\affiliation{Centre for Theoretical Physics, Polish Academy of Sciences, Al. Lotnik\'ow 32/46, 02-668 Warsaw, Poland}

\author{M. J. I. Brown\textsuperscript{\orcidlink{0000-0002-1207-9137}}}
\affiliation{School of Physics, Monash University, Clayton, VIC 3800, Australia}

\author{D. Leahy\textsuperscript{\orcidlink{0000-0002-4814-958X}}}
\affiliation{Department of Physics and Astronomy, University of Calgary, Calgary, AB T2N 1N4, Canada}

\author{A. Mailvaganam\textsuperscript{\orcidlink{0009-0003-1221-1630}}}
\affiliation{School of Mathematical and Physical Sciences, Macquarie University, Sydney, NSW 2109, Australia}
\alsoaffiliation{Macquarie University Astrophysics and Space Technologies Research Centre, Sydney, NSW 2109, Australia}

\author{J. R. Marvil\textsuperscript{\orcidlink{0000-0003-1111-8066}}}
\affiliation{National Radio Astronomy Observatory, P.O. Box O, Socorro, NM 87801, USA}

\author{T. Mukherjee\textsuperscript{\orcidlink{0009-0004-7639-869X}}}
\affiliation{School of Mathematical and Physical Sciences, Macquarie University, Sydney, NSW 2109, Australia}
\alsoaffiliation{Macquarie University Astrophysics and Space Technologies Research Centre, Sydney, NSW 2109, Australia}

\author{S. F. Rahman\textsuperscript{\orcidlink{0000-0001-9414-175X}}}
\affiliation{Lahore University of Management Sciences (LUMS), Lahore, Pakistan}
\alsoaffiliation{NCBC at NED University of Engineering and Technology, Karachi, Pakistan}

\author{T. Vernstrom\textsuperscript{\orcidlink{0000-0001-7093-3875}}}
\affiliation{Australia Telescope National Facility, CSIRO, Space and Astronomy, PO Box 1130, Bentley WA 6102, Australia}
\alsoaffiliation{International Centre for Radio Astronomy Research, University of Western Australia, 7 Fairway, Crawley, WA 6009, Australia}

\author{J. Willingham\textsuperscript{\orcidlink{0009-0001-5653-9481}}}
\affiliation{School of Mathematical and Physical Sciences, Macquarie University, Sydney, NSW 2109, Australia}
\alsoaffiliation{Macquarie University Astrophysics and Space Technologies Research Centre, Sydney, NSW 2109, Australia}

\author{T. Zafar\textsuperscript{\orcidlink{0000-0003-3935-7018}}}
\affiliation{School of Mathematical and Physical Sciences, Macquarie University, Sydney, NSW 2109, Australia}
\alsoaffiliation{Macquarie University Astrophysics and Space Technologies Research Centre, Sydney, NSW 2109, Australia}
\received {dd Mmm YYYY}
\revised  {dd Mmm YYYY}
\accepted {dd Mmm YYYY}
\published{dd Mmm YYYY}

 %% First letter not capped
\begin{document}

\begin{abstract}
While it is well known that galaxies are composites of many emission processes, quantifying the various contributions remains challenging. In this work, we use unsupervised machine learning based clustering algorithms to evaluate the agreement between the clustering tools and astrophysical classifications, and hence quantify the fractional contributions of star formation processes and nuclear black hole activity to the total galaxy energy budget of radio sources. We perform clustering on the multiwavelength (optical, infrared (IR), and radio) active galactic nuclei (AGN) diagnostic spaces, using the data from the G09 and G23 fields from the Galaxy and Mass Assembly (GAMA) survey, Evolutionary Map of the Universe (EMU) survey, and the Wide-field Infrared Survey Explorer (WISE). We find that the statistical clustering recovers $\approx$\,90\% of the star forming galaxies (SFGs) and $\approx$\,80\% of the AGN. We define a new IR-radio AGN diagnostic scheme that identifies radio AGN from IR SFGs and AGN, corresponding to the KMeans cluster with approximately 90\% reliability. We demonstrate the superior power of radio AGN selection in higher dimensions using a three-dimensional space composed of directly observable parameters ($\rm W_1-W_2$ colour, $\rm W_2$ magnitude, and the 1.4\,GHz radio flux density). This novel three dimensional diagnostic shows immense potential in radio AGN selection that is close to 90\% reliable and 90\% complete. We also publish a catalogue of radio sources in the EMU survey with associated probabilities for them to be active in the optical regime, through which we emphasise the philosophy of considering a galaxy to be composed of various fractions rather than a binary classification of SFGs and AGN.
\end{abstract}
\section{Introduction}
\label{sec:int}
Star formation and black hole accretion are two of the most prominent energy sources that contribute to the emission from a galaxy. The interplay between these two processes is driven by an observed co-evolution of black holes and their host galaxies \citep{2013ARA&A..51..511K}, such that these processes either boost or quench each other. Signatures of these processes are ubiquitous in the spectra and colours of galaxies. Disentangling them is crucial in understanding the evolutionary properties of galaxies and in extracting important astrophysical parameters (e.g. the star formation rate (SFR) and stellar mass). Traditionally, this has been achieved using optical spectral line ratios \citep[the BPT diagram and its variants;][]{1981PASP...93....5B,1985PASP...97.1129O,1987ApJS...63..295V}, infrared (IR) colours \citep[e.g.][]{2012ApJ...753...30S,2012ApJ...754..120M,2018ApJS..234...23A}, radio luminosity threshold \citep[e.g.][]{2000ApJS..126..133W,2017A&A...602A...3D,2017MNRAS.464.3271M,2025PASA...42...77P}, X-ray luminosity threshold \citep[e.g.][]{2022PASJ...74..689F,2023MNRAS.523.4756B}, spectral energy distributions \citep[SEDs, e.g.][]{2018MNRAS.473.3710C,2024PASA...41...16P}, and various other multiwavelength diagnostic techniques.

The energy output of galaxies is not always dominated by star formation or black hole accretion alone (quiescent and passive galaxies are some examples); although in many cases, both contribute significantly. Traditional galaxy classification techniques approach this problem in a binary perspective, thereby classifying galaxies into one of two classes: star forming galaxies (SFG) and active galactic nuclei (AGN). A problem with this approach is that the exact astrophysics of the emission mechanisms gets hidden underneath the binary label. This can be solved by quantifying the fractional contributions from the various processes in a galaxy and is readily achieved by SED analysis, integral field studies \citep[e.g.][]{2024ApJ...971..165X}, and by radio observations accurately localising the radio cores \citep[e.g.][]{2025MNRAS.536L..32M}. There have been attempts to quantify these contributions in terms of probabilities by identifying regions in the traditional diagnostic diagrams where there is potential mixing between the populations \citep[e.g.][]{2011ApJ...736..104J,2013ApJ...764..176J,2011ApJ...728...38Y,2018ApJ...861L...2T,2023MNRAS.523.4756B,2025MNRAS.536L..32M}.

This work investigates whether a binary classification of galaxies as either star forming or AGN-dominated is appropriate from a statistical viewpoint using clustering analysis, with a focus on radio detections. Clustering is an unsupervised machine learning (ML) technique, a task of grouping similar objects by some measure, for instance, Euclidean distance between the data points. Broadly, existing clustering techniques can be divided into two categories: hard clustering, where each data point completely belongs to a given cluster, and soft clustering, where likelihoods are evaluated for the data points and clusters to which they belong \citep{2022MNRAS.517.5496Y}. The clustering tools are also divided into different types, based on the employed principles, such as partition-based (e.g. KMeans; \citealt{mcqueen1967smc}, fuzzy-c-means, FCM; \citealt{BEZDEK1984191}), density-based (e.g. density-based spatial clustering of applications with noise, DBSCAN; \citealt{1996kddm.conf..226E}), hierarchical \citep[e.g. Balanced Iterative Reducing and Clustering Using Hierarchies, BIRCH;][]{Zhang1996BIRCHAE}, statistical model-based \citep[Gaussian mixture models, GMM;][]{everitt2011cluster}, and grid-based.

ML tools have been applied in various multidimensional studies, including galaxy classification in high-dimensional parameter spaces \citep[e.g.][]{2003PASP..115.1006Z,2004A&A...422.1113Z}, for star-galaxy-quasar separation \citep[e.g.][]{2020A&A...633A.154L,2021A&A...645A..87B}, and AGN/SFG classification \citep[e.g.][]{2022MNRAS.515.6046P,2023A&A...675A.159K,2025MNRAS.537.1028P}. In this work, we evaluate the performance of four unsupervised clustering algorithms: KMeans, GMM, FCM, and BIRCH, on multiwavelength AGN diagnostic diagrams. We choose these algorithms from a pool of clustering tools since they produced results comparable with the AGN classification. Clustering is performed in a multidimensional parameter space defined by features from the multiwavelength AGN diagnostics, and the resulting clusters are examined through their projections onto widely used two-dimensional diagnostic tools. These include optical diagnostics such as the BPT diagram \citep{1981PASP...93....5B}, the mass–excitation (MEx) diagram \citep{2011ApJ...736..104J,2013ApJ...764..176J}, the blue diagram \citep{2010A&A...509A..53L}, and the colour-excitation (CEx) diagram \citep{2011ApJ...728...38Y}; IR diagnostics such as the WISE\footnote{Wide-field Infrared Survey Explorer} colour-based diagrams \citep{2012MNRAS.426.3271M,2018ApJS..234...23A} and the KI diagram \citep{2012ApJ...754..120M,2014A&A...562A.144M}; and the MId-infrared and RADio diagram \citep[MIRAD;][]{2021ApJ...910...64K}.

The structure of the paper is as follows: \S\,\ref{sec:data} describes the multiwavelength dataset used in this work. \S\,\ref{sec:agn_classification} introduces the various AGN diagnostic tools and \S\,\ref{sec:clustering} discusses the procedure followed in clustering these diagnostic spaces. \S\,\ref{sec:results} presents the results, where we also introduce a new IR-radio AGN diagnostic tool. We discuss the findings in \S\,\ref{sec:discussion} and conclude in \S\,\ref{sec:conclusion}. The optical magnitudes used in this work are in the AB system, and the WISE magnitudes are in the Vega system. The WISE magnitudes have been converted to AB magnitudes only in the instances where they are combined with the K-band magnitudes (the \citealt{2012ApJ...754..120M} diagnostic, \S\,\ref{sec:ir_agn}). Throughout the paper we assume a flat $\Lambda$CDM cosmology with the  parameters $H_o=70\,$km\,s$^{-1}$\,Mpc$^{-1}$, $\Omega_M=0.3$, and $\Omega_\Lambda=0.7$.
\section{Data}\label{sec:data}
The data required for AGN diagnostics and associated clustering come from various multiwavelength surveys: Galaxy and Mass Assembly \citep[GAMA;][]{2011MNRAS.413..971D,2022MNRAS.513..439D} survey for optical photometry and spectra, Evolutionary Map of the Universe \citep[EMU\footnote{\hyperlink{https://emu-survey.org/}{https://emu-survey.org/}};][]{2011PASA...28..215N,2021PASA...38...46N,2025PASA...42...71H} survey for the radio data, and WISE survey \citep{2010AJ....140.1868W} for the IR magnitudes.
\subsection{Optical Data}\label{sec:optical_data}
The optical AGN diagnostics of interest in this work use spectral line intensities and photometric magnitudes, which are obtained from the GAMA survey. GAMA is a multiband imaging and spectroscopic survey undertaken at the Anglo-Australian Telescope (AAT) over five separate fields (G02, G09, G12, G15, and G23) with a limiting magnitude of $m_r<19.8$ ($m_i<19.2$ for the G23 field), totalling 286 deg$^2$. The GAMA fields other than G02 benefit from the overlap with multiple multiwavelength surveys \citep{2020MNRAS.496.3235B}, such as the Galaxy Evolution Explorer \cite[GALEX,][]{2005ApJ...619L...1M}, European Southern Observatory (ESO) VLT Survey Telescope Kilo Degree survey \citep[KiDS,][]{2015A&A...582A..62D}, ESO Visible and Infrared Survey Telescope for Astronomy (VISTA) Kilo-degree Infrared Galaxy Public survey \citep[VIKING,][]{2013Msngr.154...32E}, WISE, and Herschel \citep{2010A&A...518L...1P} Astrophysical Terahertz Large Area Survey (ATLAS) \citep{2010PASP..122..499E}. Along with the multi-band photometry, these fields also benefit from the spectroscopic data from the AAT AAOmega spectrograph. We are interested in the G09 (129$\rm ^o$<RA<141$\rm ^o$, -2$\rm ^o$<$\delta$<3$\rm ^o$) and G23 (339$\rm ^o$<RA<351$\rm ^o$, -35$\rm ^o$<$\delta$<-30$\rm^o$) fields since these fields are covered by the EMU survey (see \S\,\ref{sec:emu_data} below). Both of these fields also benefit from a high spectroscopic completeness of 98.5\% in the G09 field and 94.2\% in the G23 field \citep{2015MNRAS.452.2087L}.

GAMA data is arranged as tables in data management units (DMUs). The \verb|gkvScienceCatv02| \citep{2020MNRAS.496.3235B} table in the \verb|gkvInputCatv02| DMU contains the optical magnitudes, which also provides the necessary $\rm K$-band magnitudes (see \S\,\ref{sec:final_sample} and \S\,\ref{sec:ir_agn}). Stellar mass estimates are adopted from the \verb|StellarMassesGKVv24| table \citep{2011MNRAS.418.1587T} in the \verb|StellarMassesv24| DMU. The spectral lines ([N{\sc ii}], [O{\sc ii}], [O{\sc iii}], H$\alpha$, and H$\beta$) are taken from the \verb|GaussFitSimplev05| table \citep{2017MNRAS.465.2671G} in the \verb|SpecLineSFRv05| DMU. We restrict to sources flagged \verb|IS_BEST|, indicating reliable spectroscopic measurements. This criterion makes sure that we select the best spectrum for a given source from GAMA itself since the catalogue contains spectra from other surveys as well for comparison \citep[see][for more details]{2017MNRAS.465.2671G}. The G09 and G23 fields contain 58\,498 and 44\,476 optical sources, respectively. This implies a total of 102\,974 sources with optical spectra, photometry and stellar masses. Additional signal-to-noise (S/N) cut-off applied on the spectral lines for the optical sample is described in \S\,\ref{sec:final_sample}.
\subsection{Radio flux densities}\label{sec:emu_data}
EMU \citep{2011PASA...28..215N,2021PASA...38...46N,2025PASA...42...71H} is an ongoing wide-field radio continuum survey project to deliver the touchstone radio atlas of the southern hemisphere using the Australian Square Kilometre Array Pathfinder \citep[ASKAP;][]{2021PASA...38....9H} radio telescope.%Its primary goal was a 3$\pi$ sr sky survey at 943\,MHz up to $\delta=+30^o$ reaching a root mean square (rms) noise of $\sigma=20-30\,\mu\,\rm Jy\,beam^{-1}$ with a resolution $\sim15''$. 
The survey aims at covering $2\pi$ sr of the sky and cataloguing 20 million extragalactic sources, including SFGs up to redshift $z\sim1$, powerful starbursts and AGN reaching even greater redshifts \citep{2025PASA...42...71H}.

Our focus is on the two GAMA fields, G09 and G23, for the integrated radio flux densities. The G23 field has been observed as a part of the EMU early science project \citep{2022MNRAS.512.6104G} covering an area of 82.7 deg$^2$ with a sensitivity of $\rm 0.038\,mJy\,beam^{-1}$ at a frequency of 887.5\,MHz. Out of the 55\,247 sources detected, 39\,812 have S/N ratios $\geq$ 5. The G09 field has been observed as a part of the EMU main project at 943\,MHz and the data is internally released. The G09 field as processed by the EMUCAT pipeline \citep[][Marvil et al., in prep.]{2025PASA...42...71H} covers an area of around 96 deg$^2$ with 58\,898 sources, where 37\,981 of them have S/N$\geq$5. Since the radio diagnostic described in \S\,\ref{sec:radio_agn} uses the radio flux density at 1.4\,GHz (\radioflux), the radio flux densities of both G09 and G23 sources are scaled to 1.4\,GHz assuming synchrotron emission with a spectral index $\alpha=-\,0.7$, where $S_\nu=\nu^\alpha$. We do not analyse extended and compact radio sources separately. An analysis with this focus is being done by Barnes et al. (in prep), where extended radio sources are identified and classified using ML tools.
\subsection{Infrared magnitudes}\label{sec:wise_data}
WISE is an all sky IR survey that scanned the entire sky in four bands, namely $\rm W_1, W_2, W_3$, and $\rm W_4$ corresponding to 3.4, 4.6, 12, and 22 microns with 5$\sigma$ sensitivities of 0.08, 0.11, 0.8, and 6 mJy, respectively. The $\rm W_1, W_2$, and $\rm W_3$ bands have better sensitivities and S/N values compared to $\rm W_4$ \citep{2010AJ....140.1868W}, due to which we focus on the first three bands. Subsequent data releases of the WISE survey include the AllWISE \citep{2012yCat.2311....0C,2014yCat.2328....0C} catalogue consisting of 747\,634\,026 sources with the four magnitudes. The latest data release, the CatWISE2020 catalogue \citep{2021ApJS..253....8M} contains 1\,890\,715\,640 sources, but with only the $\rm W_1$ and $\rm W_2$ magnitudes. Since the diagnostic tools that we are interested in this work require the three WISE bands, we choose the AllWISE catalogue for IR magnitudes (profile-fit photometry). The three WISE bands used for analysis (see \S\,\ref{sec:ir_clustering}) have a cut-off of S/N$\geq$5 applied (see \S\,\ref{sec:final_sample} for details). The CatWISE2020 catalogue is used only for positional crossmatching as an intermediate step in identifying radio and optical counterparts (see \S\,\ref{sec:final_sample}), not for IR magnitudes. Accordingly, some radio sources with optical counterparts in GAMA may be lacking AllWISE counterparts and their IR magnitudes.
\subsection{Final samples}\label{sec:final_sample}
We use two separate samples for clustering in optical, IR, and radio regimes. Combining all three datasets would exclude many IR and radio AGN \citep{2019PASA...36...24L} since an AGN may not show AGN signatures in all of these bands simultaneously \citep{2014ARA&A..52..589H}. A solution would be using a censored dataset, where for some galaxies, certain physical attributes are missing. For instance, optical-IR-radio catalogues will have galaxies detected in IR and radio, where optical counterparts do not exist. We could model the parameters of interest for such a sample in the clustering space by masking the missing parameters. This does not mean data imputation, a statistically driven approach where the values of the missing parameters are manually inserted, because such a procedure can lead to a bias towards the properties of the sample at hand. Rather, a data censoring approach would involve constructing a marginal probability distribution without the missing parameter so as to enable clustering. However, at this stage, this procedure was deemed beyond the scope of this work (see \citealt{2025MNRAS.537.1028P} and Carvajal et al., in prep for similar approaches), so we decided to analyse the optical and IR-radio regimes separately with an optical-radio sample and an IR-radio sample, respectively, keeping radio data as the link between the two spectral regimes.
%For crossmatching between the optical and the radio catalogues however, we use the CatWISE positions (we associate radio sources with their corresponding IR counterparts and match the IR positions with the optical positions), since they are more accurate with respect to AllWISE \citep{2021ApJS..253....8M}. The crossmatching procedure described here is applicable to both G09 and G23 tables (column names of different parameters could be different), and we collectively label the optical sample as `GAMA'. The IR catalogue with CatWISE positions and AllWISE magnitudes is labelled `WISE', and the radio sample as `EMU'. We used Topcat \citep{2005ASPC..347...29T} for crossmatching the different catalogues in this work.

Radio sources are often cross-matched with the optical catalogue using their IR counterparts (\citealt{2021PASA...38...46N}, also see Anih et al., in prep for a similar strategy selecting near-IR counterparts to X-ray sources). This is because of the strong correlation between the IR and radio emission in galaxies \citep{1992ARA&A..30..575C}. Also, WISE provides precise positions with respect to EMU because of its better spatial resolution. We follow this approach and as the first step in generating the optical-radio sample, we crossmatch the radio data with the CatWISE2020 catalogue (to get the coordinates of the IR counterparts of the radio sources)  with a 5" radius \citep{2021PASA...38...46N}, resulting in 37\,981 sources in G09 and 34\,841 sources in G23. CatWISE2020 catalogue incorporates a much longer time baseline relative to AllWISE, resulting in, although similar positional accuracies, but likely includes fainter sources \citep{2021ApJS..253....8M}. Because of this, we use the CatWISE2020 positions instead of AllWISE for cross-matching. This radio catalogue containing the CatWISE2020 positions is then cross-matched with GAMA with a radius of 5" \citep{2021PASA...38...46N}, resulting in 5\,119 G09 sources and 5\,890 G23 sources. WISE sources have an angular resolution of 6" in the first three bands and 12" in $\rm W_4$ \citep{2010AJ....140.1868W}. EMU has an angular resolution of 15" \citep{2025PASA...42...71H} which, when compared with the WISE resolution, deems 5" a fair cross-match radius. Some radio sources are extended objects, which can result in spurious matches when a fixed crossmatch radius is used, especially when finding optical counterparts. The approach we adopted here, using IR positions for finding optical matches and the choice of an appropriate crossmatch radius \citep{2024PASA...41...21A}, helps in minimising spurious matches.
%GAMA already provides WISE flux densities, but since we also have the IR-radio sample with no GAMA measurements, we use the AllWISE magnitudes for the GAMA sources for uniformity.
%As the first step, EMU and WISE are cross-matched with a radius of 5", resulting in 37\,981 sources in G09 and 34\,841 sources in G23. This EMU-WISE catalogue is then crossmatched with GAMA with a radius of 5", but with IR (CatWISE) coordinates of the radio sources. This results in 5\,119 G09 optical sources and 5\,890 G23 optical sources.

\begin{figure}[!hbt]
    \centering
    \includegraphics[width=\linewidth]{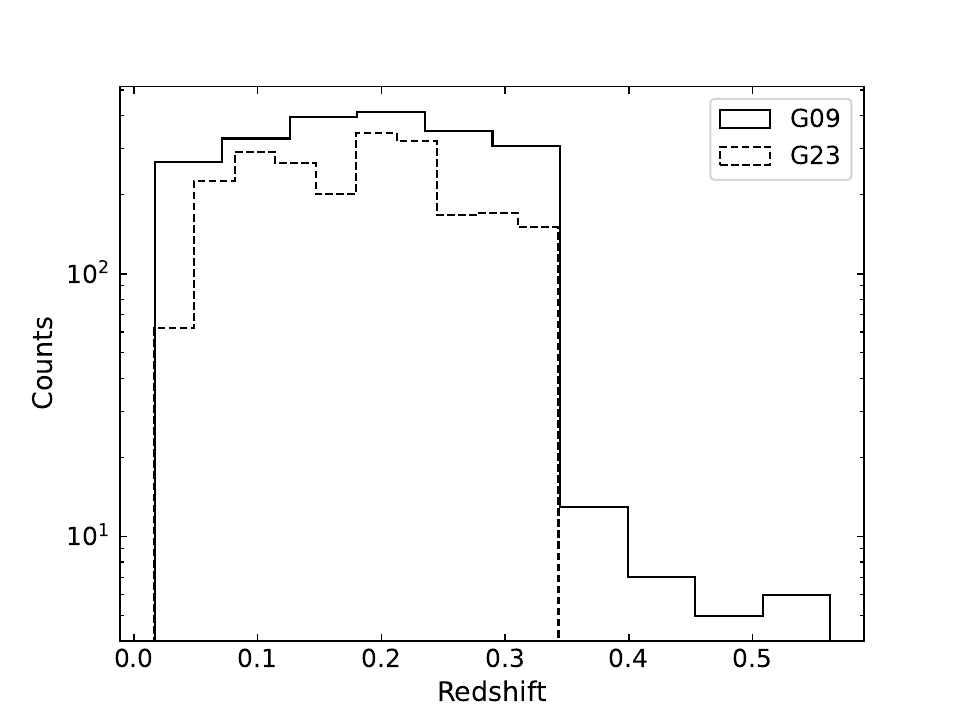}
    \caption{The redshift distributions of the radio sources in G09 (solid line) and G23 (dashed line) fields. These radio sources belonging to the optical-radio sample are selected as described in the text (see \S\,\ref{sec:final_sample}). The requirement of H$\alpha$ line for BPT classification results in the sudden drop at $z\approx0.34$. A few higher redshift G09 objects are most likely resulting from the higher completeness and slightly fainter magnitude limit of the field.}
    \label{fig:redshift}
\end{figure}

We applied quality cuts such that the various spectral line ratios used in the optical diagnostic tools (\logniiha,\,\logoiihb,\,and \logoiiihb) have S/N$\geq$1. We expect this S/N cut to remove the radio AGN with passive spectra or undetected lines. The effects of this selection bias are discussed in \S\,\ref{sec:cluster_disc}. Certain spectral line values during the fitting procedure are replaced by the dummy value `-99999.0' when there are not enough pixels to perform a fit due to bad pixels or limits of the spectral wavelength range \citep{2017MNRAS.465.2671G}. After removing such entries from the line ratios, the final sample contains 2\,091 G09 optical-radio sources and 2\,193 G23 optical-radio sources. Figure\,\ref{fig:redshift} shows the redshift distributions of these samples, where solid and dashed lines correspond to G09 and G23 sources, respectively. The sudden drop in the redshift distribution of both G09 and G23 sources can be attributed to the requirement of H$\alpha$ emission lines for the optical AGN classification (see \S\,\ref{sec:optical_agn}). Few objects in the G09 sample extend to higher redshifts, most likely resulting from its higher completeness and slightly fainter magnitude limit. We do not apply a higher S/N, for instance, a $\rm S/N=5$ threshold, on the spectral lines because this will drastically reduce the sample size. Additionally, the results described in \S\,\ref{sec:results} are not strongly dependent on this choice.
\begin{table}[!htb]
    \centering
    \begin{tabular}{c|c|c}
    \hline
       Field & Optical-radio & Infrared-radio \\
       \hline
        G09  & 2\,091        & 14\,123 \\
        G23  & 2\,193        & 14\,339 \\
        \hline
        Total & 4\,284        & 28\,462
    \end{tabular}
    \caption{The number of sources in the final samples used in this work. Both optical and IR catalogues are crossmatched with radio sources as described in \S\,\ref{sec:final_sample}.}
    \label{tab:final_sample}
\end{table}

For the IR-radio sample, we crossmatch the AllWISE and radio catalogues with a radius of 5" \citep{2021PASA...38...46N}. We use the AllWISE catalogue here (and not the CatWISE2020 positions as we did for optical-radio matching) because the IR diagnostic tools employed in this work require photometry in the first three WISE bands (see \S\,\ref{sec:ir_clustering}). We use the same cross-match radii while using CatWISE2020 or AllWISE catalogues because they have similar positional accuracies \citep{2010AJ....140.1868W,2021ApJS..253....8M}. This crossmatch resulted in 32\,844 G09 sources and 28\,947 G23 sources. We applied a quality cut on the three ($\rm W_1,\,W_2$, and $\rm W_3$) WISE bands so that the magnitudes satisfy S/N$\geq$5, resulting in 14\,123 G09 IR-radio sources and 14\,339 G23 IR-radio sources. It should be noted that the KI diagram described below in \S\,\ref{sec:ir_agn} requires having a K-band magnitude. This restricts those results to only a much smaller number of sources (3755 in G09 and 4101 in G23). Because of this, we do not include the K-band magnitudes as an additional parameter in the clustering analysis. We also do not apply any S/N criterion on K-band magnitudes.
\section{AGN classification}
\label{sec:agn_classification}
We employ a suite of multiwavelength spectral and photometric AGN diagnostics in this work, comprising four optical diagnostics, three IR colour-magnitude diagnostics, and one IR-radio flux-based diagnostic. We do not employ X-ray data in this analysis, as a crossmatch of the relevant samples with the X-ray data \citep[first eROSITA all-sky survey (eRASS1);][]{2024A&A...682A..34M} resulted in a sample of only a few hundred galaxies, which would preclude the clustering analysis. This section details the relevant diagnostic tools and our approach in employing them for our purposes. These diagnostics are visualised in Figures\,\ref{fig:Optical_clusters} and \ref{fig:ir_radio_clusters}. The sources in the respective samples are classified using these diagnostics in a binary fashion, and these labels are used only for comparison with the clustering outputs described in \S\,\ref{sec:clustering}.
\subsection{Optical diagnostics}\label{sec:optical_agn}
\subsubsection{BPT diagram}\label{sec:bpt}
%The Baldwin, Phillips, Terlevich (BPT) diagram saw its inception in the work done by \cite{1981PASP...93....5B}. \cite{1971ApJ...168..327S} had already demonstrated that HII regions can be classified to distinguish their potential ionising sources based on the strength of their emission lines and their ratios. In its complete form,
The Baldwin, Phillips, and Terlevich (BPT) diagram saw its inception in the work done by \cite{1981PASP...93....5B}. The BPT diagram uses the spectral line ratios \logniiha\,and \logoiiihb\,for ionising source classification into AGN and SFGs, and is used extensively \citep[e.g.][]{2024PASA...41...16P,2025PASA...42...77P,2025arXiv250914019H,2025arXiv250906913S}. Further improvements were proposed by \cite{2001ApJS..132...37K} for a theoretical maximum starburst line and by \cite{2003MNRAS.346.1055K} for a semi-empirical classification \citep[also see,][]{1985PASP...97.1129O,1987ApJS...63..295V}. The BPT diagram, as used in this work (see Figure\,\ref{fig:Optical_clusters}a), classifies galaxies into three categories: $(i)$ SFGs, if they fall below the \cite{2003MNRAS.346.1055K} line, $(ii)$ AGN, if they lie above the \cite{2001ApJS..132...37K} line, and $(iii)$ composites, if they fall in between the two demarcation lines.
\subsubsection{MEx diagram}\label{sec:mex}
The mass-excitation (MEx) diagram was proposed by \cite{2011ApJ...736..104J} as an alternative to traditional nebular line diagnostics as the [N{\sc ii}] and H$\alpha$ lines are redshifted out of the optical spectra for sources at high redshifts (say, for $z>\approx0.5$). The MEx diagram uses \logoiiihb\, and galaxy stellar mass as a proxy for \logniiha\,. The choice of stellar mass instead of \logniiha\, can be justified using the mass-metallicity relationship \citep{2004ApJ...613..898T}. The MEx diagram classifies galaxies into SFGs, AGN, and composites (see Figure\,\ref{fig:Optical_clusters}b). In this work, we use the demarcation lines defined in \cite{2013ApJ...764..176J}.
\subsubsection{Blue diagram}\label{sec:blue}
Galaxy classification has also been explored using only the blue spectral lines ([O{\sc iii}], [O{\sc ii}], and H$\beta$) to enable the analysis of high redshift sources. \cite{2004MNRAS.350..396L} introduced the blue diagram which uses the line ratios \logoiihb\,and \logoiiihb\,and was improved by \cite{2010A&A...509A..53L}. The diagram contains five regions corresponding to SFGs, AGN, low-ionisation nuclear emission line regions (LINERs), composites of SFGs and AGN, and the composites of SFGs and LINERs. The true nature of LINERs is actively debated where spatially resolved studies \citep[e.g.][]{2016MNRAS.461.3111B} have identified LINER-like spectra in star-forming and quenched galaxies. \cite{2017FrASS...4...34M} demonstrated that a large fraction of the LINERs in their sample of X-ray sources can be considered genuine AGN. In this work, we classify LINERs as AGN, thus reducing the composite class to a single population, resulting in a three-system classification like the BPT and MEx classifications (see Figure\,\ref{fig:Optical_clusters}c).
\subsubsection{CEx diagram}\label{sec:cex}
The colour-excitation (CEx) diagram  \citep{2011ApJ...728...38Y} follows a similar approach to the MEx and the blue diagrams, to be able to classify high redshift galaxies. The CEx diagram is inspired by the correlation between rest frame $\rm H$-band magnitude and the metallicity for SFGs \citep{2007ApJ...660L..39W,2011ApJ...728...38Y}. This relation collapses for AGN and thus the $\rm U-B$ colour, along with \logoiiihb, was proposed as a more effective probe for galaxy classification. The CEx diagram classifies galaxies into two classes, SFGs and AGN; a composite region is not defined in this case (see Figure\,\ref{fig:Optical_clusters}d). We used the K-correction relations in  \cite{2007AJ....133..734B} to match the $\rm U-B$ criterion of \cite{2011ApJ...728...38Y} to the $\rm u-g$ colour generated using GAMA data.
\subsection{Infrared diagnostics}\label{sec:ir_agn}
AGN and star formers show significant differences in their IR SEDs \citep[e.g.][]{2022Univ....8..304L}. This feature has been exploited in the form of near- and mid-IR colour criteria in developing IR AGN diagnostic diagrams. Multiple surveys have covered the sky in IR. These include the Two-Micron All Sky Survey \citep[2MASS;][]{2006AJ....131.1163S}, Spitzer Space Telescope \citep{2004ApJS..154....1W} using the infrared array camera \citep[IRAC;][]{2004ApJS..154...10F}, and WISE. The AGN selection criteria developed using the data from these surveys include, the Lacy wedge \citep{2004ApJS..154..166L, 2007AJ....133..186L, 2005ApJ...621..256S}, the Stern wedge \citep{2005ApJ...631..163S}, the IRAC power-law \citep{2006ApJ...640..167A, 2007ApJ...660..167D, 2012ApJ...748..142D}, the KI \& KIM methods \citep{2012ApJ...754..120M, 2014A&A...562A.144M}, and the WISE selections \citep{2012ApJ...753...30S,2013ApJ...772...26A,2018ApJS..234...23A}. In this work, we focus on the three diagnostic tools that can be built using WISE magnitudes.

\cite{2012ApJ...753...30S} put forward a magnitude-independent AGN selection criterion that selects AGN reliably up to $\rm W_2$ magnitude brighter than 15.05 mag. As an improvement, \cite{2018ApJS..234...23A} proposed a magnitude-dependent selection, with $\rm W_1-W_2$ colour as a function of $\rm W_2$ magnitude, that selects AGN reliably up to $\rm W_2$ magnitudes brighter than 17.11 mag. The demarcation line in this diagram varies depending on the reliability and completeness of the resulting AGN sample. In this work, we use the version optimised for reliability \citep[equation 4 of][]{2018ApJS..234...23A}, that classifies the galaxies into SFGs and AGN (see Figure\,\ref{fig:ir_radio_clusters}a).

\cite{2012MNRAS.426.3271M} defined a wedge that primarily selects X-ray AGN that are also active in the IR. They proposed two versions of the diagram, where the first version uses the first three WISE magnitudes, and the second uses all four WISE magnitudes. We choose the former in this work, because the $\rm W_4$ band is relatively shallow and may only select bright objects. The three-band wedge uses the colours $\rm W_1-W_2$ and $\rm W_2-W_3$ for delineating the SFGs and AGN (see Figure\,\ref{fig:ir_radio_clusters}b).

The KI diagram proposed by \cite{2012ApJ...754..120M} uses the K and IRAC bands (K-IRAC colour). IRAC bands facilitate AGN identification since they capture the polycyclic aromatic hydrocarbon (PAH) features in the mid-IR bands. In the EMU fields considered in this work, however, we do not have IRAC data available. Instead, we replace the IRAC bands in the KI diagram with the WISE bands. The 4.5 micron IRAC band is close to the $\rm W_2$ band (4.6 microns) in wavelength. Instead of the 8 micron IRAC band, we use the WISE $\rm W_3$ band (12 microns). Both WISE bands are converted to AB magnitudes in this case since the \cite{2012ApJ...754..120M} criteria are defined in the AB system. We acknowledge the differences in the widths of the IRAC and WISE bands, and that they might probe additional features than what the corresponding IRAC bands do. As evident in \S\,\ref{sec:results}, this difference in wavelength does not drastically affect the classification. The \cite{2012ApJ...754..120M} diagnostic also classifies the galaxies into SFGs and AGN. There are no composite classes identified in the IR diagnostics used in this work (see Figure\,\ref{fig:ir_radio_clusters}c).
\subsection{Radio diagnostics: The MIRAD diagram}\label{sec:radio_agn}
Galaxies with radio emission driven by black hole accretion are termed radio AGN. Various classes of such objects exist, and multiple definitions based on specific use cases are employed while selecting radio AGN \citep[see][for a review on AGN and different types of radio AGN]{2014ARA&A..52..589H}. The most prominent radio AGN selection uses the FIR–radio correlation \citep{1992ARA&A..30..575C}, a relation tightly followed by SFGs. Sources that deviate significantly from this correlation are interpreted as radio loud AGN. An upper limit on the radio luminosity \citep{2000ApJS..126..133W,2025PASA...42...77P}, and radio-to-optical flux ratio \citep{1989AJ.....98.1195K} are also widely used techniques for radio loud AGN selection, hereafter, termed radio AGN. As discussed in \cite{2025PASA...42...77P}, the scarcity of ample radio sources with spectroscopic counterparts restricts the usage of luminosity as a classification criterion in many cases. An alternative to this approach is the flux-flux diagrams with an appropriate selection criterion (similar to flux-ratio methods). 

Given these considerations, in this work we use the MId-infrared and RADio \citep[MIRAD;][]{2021ApJ...910...64K} flux-based diagram for radio AGN selection (see Figure\,\ref{fig:ir_radio_clusters}d). It uses the WISE $\rm W_3$ flux, \radioflux, and a simple one-to-one line as a selection criterion. In this diagram, pure star formers occupy the upper sequence and AGN, since they generally show excess radio emission for a given mid-IR flux, occupy the lower sequence. It is interesting to note that recent studies using LOw-Frequency ARray (LOFAR) \citep[e.g.][]{2025MNRAS.536L..32M} have shown that many compact sources without excess radio emission also host AGN-related radio emission. In these perspectives, this selection is potentially a lower limit on the radio AGN numbers. % Using fluxes instead of luminosities introduce additional complexities due to distance dependence and results in a bias against faint sources (see \S\,\ref{sec:ir_clustering} and Figure\,\ref{fig:ir_radio_clusters}d). 
\section{Clustering}\label{sec:clustering}
%In this section, we discuss the suite of clustering tools used in analysing the AGN diagnostic diagrams described in \S\,\ref{sec:agn_classification}. The final optical-radio and IR-radio samples are classified using AGN diagnostics in each of these wavelengths, resulting in AGN/SFG labels for every source where possible. In the analyses, however, we do not use these classifications, instead we are interested in the labels assigned to the sources by the clustering algorithms.
A wide range of clustering algorithms is available, and assessing their performance in each diagnostic space is essential. In this work, we use the KMeans, GMM, FCM, and BIRCH algorithms to statistically classify the radio galaxies. These four tools are chosen since they represent different clustering principles, so that the performance of both the clustering principles (see the subsections below for a description of the algorithms and the principles employed by them) and the tools can be assessed. The input to the clustering tools is a set of physical observables used in the diagnostic diagrams discussed earlier, resulting in an $n$ dimensional space, where the clustering is realised. Analyses in higher dimensional space, as we encounter here, can be supported by dimensionality reduction techniques like principal component analysis \citep[PCA;][]{10.1093/biomet/28.3-4.321} where we can choose the components carrying the most variance among the whole input parameter set. Clustering can then be done on these principal components with the reduced dimensional complexity. As we show later in \S\,\ref{sec:results}, PCA is an optional step, and whether it supports the clustering is dependent on the specific spectral regime. All parameters used for clustering are standardised using the \verb|RobustScaler| method in the \verb|Python| package \verb|scikit-learn| \citep{scikit-learn} to a mean of zero and standard deviation of one. This is important, especially in astronomical data, since most parameters scale logarithmically. In the rest of this section, we give an overview of the different algorithms used.
\subsection{KMeans}\label{sec:kmeans}
KMeans belongs to the broad category of centroid (partition)-based clustering \citep{2022MNRAS.517.5496Y}. It is a simple and robust algorithm, iteratively classifying the data into a predefined number of clusters by minimising the total sum of the distances between the data points and the centres of the respective clusters \citep{2024A&C....4800851F}. KMeans is used extensively in astronomy for multiple applications, such as spectral classification and improving S/N values by stacking the spectra \citep[see,][and references therein]{2014A&A...565A..53O}. The two main decisions one has to make while initialising KMeans are $(i)$ the threshold on the distances between the data points and $(ii)$ the number of cluster centroids ($k$), which in turn decides the number of clusters. Each point is assigned to the cluster centred at the closest centroid based on the defined distance metric. This process is repeated, taking the mean positions of the points associated with a given cluster until convergence. Thus, the choices initially made can have a great impact on the outcome and the convergence time \citep{2019arXiv190407248B}.

Choosing the best $k$ value is non-trivial. It is usually done using the elbow method, which minimises the inertia, a measure of the sum of the squared distances of each data point to its closest cluster centre \citep[also see,][]{2010ApJ...714..487S}. Another measure is the silhouette coefficient, which quantifies how far away the data points are from the neighbouring cluster. We use the \verb|KMeans| function from the \verb|Python| package \verb|scikit-learn| to perform the KMeans clustering in our analysis. We used both the elbow method and the silhouette coefficient method to identify the best $k$ value, which turned out to be either three or four. As described below in \S\,\ref{sec:results}, we choose three as the value of $k$, the number of clusters, in the case of all the other clustering tools discussed.
\subsection{Gaussian mixture models}\label{sec:gmm}
GMMs are parametric probability density functions represented as a weighted sum of Gaussian probability densities. GMM, as a clustering tool, belongs to the general statistical model-based clustering methods, which models the observed data distribution as a random variable drawn from an underlying multivariate distribution \citep[][and references therein]{2017MNRAS.472.2808D,2024A&C....4800851F}.
\begin{figure*}[!htb]
    \centering
    \includegraphics[width=\linewidth]{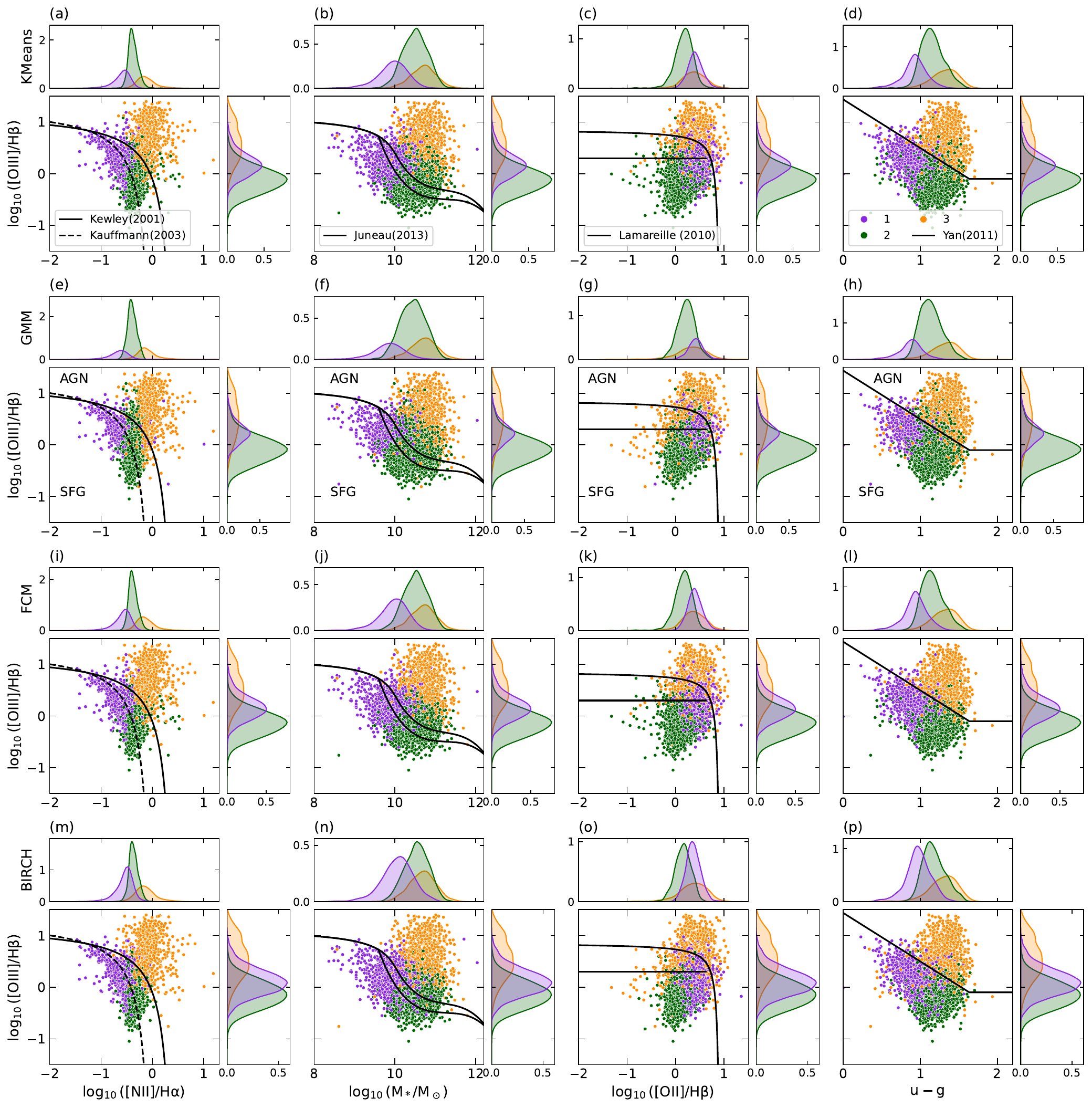}
    \caption{Performance of various clustering tools in different optical diagnostic spaces. The figure shows the different clustering tools row-wise, and the different empirical diagnostic tools are presented along the columns. Panels a-d: the clusters identified by KMeans are plotted in the optical diagnostic spaces BPT diagram, MEx diagram, blue diagram, and CEx diagram, respectively. Panels e-h: the clusters identified by GMM, panels i-l: the clusters identified by FCM, and panels m-p: the clusters identified by BIRCH are plotted in the same order as the empirical diagnostics. The SFG and AGN regions are labelled in the second row. Composite galaxies in these diagnostic plots occupy the region between the demarcation lines in the first three columns, the CEx diagram does not define a composite region. The purple and green clusters correspond to the star forming species, following a metallicity sequence. The orange clusters seemingly occupy the region identified as AGN in each of these diagnostic spaces. Each of these plots features two marginal histograms showing the normalised densities corresponding to the three clusters identified by each of the clustering tools.}
    \label{fig:Optical_clusters}
\end{figure*}
\begin{figure*}[!htb]
    \centering
    \includegraphics[width=\linewidth]{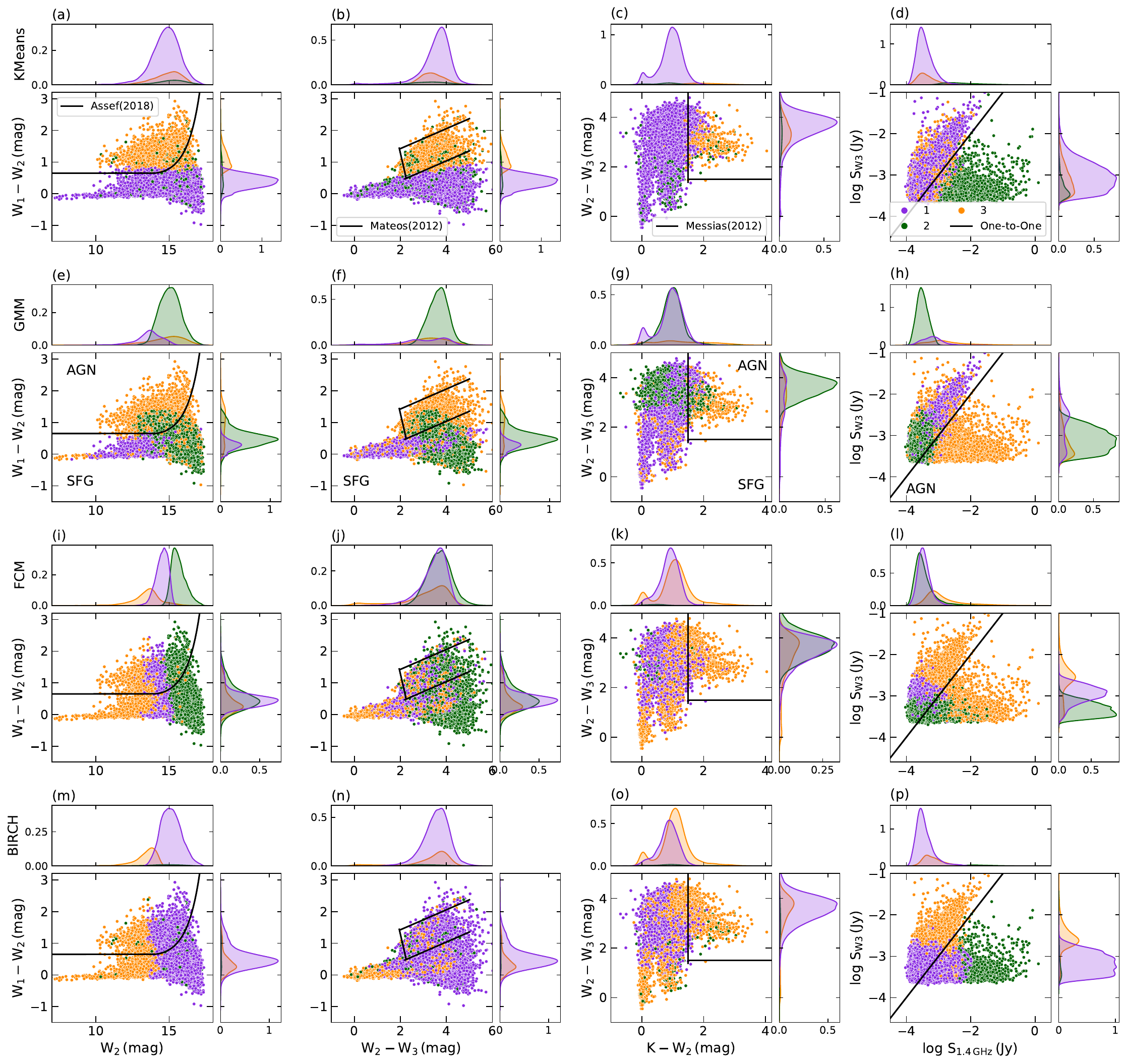}
    \caption{Performance of various clustering algorithms in different IR and radio diagnostic spaces. The figure shows the different clustering tools row-wise, and the different empirical diagnostic tools are presented along the columns. Panels a-d: the clusters identified by KMeans are plotted in the IR diagnostic spaces defined by \cite{2018ApJS..234...23A}, \cite{2012MNRAS.426.3271M}, \cite{2012ApJ...754..120M}, and the IR-radio diagnostic space defined by \cite{2021ApJ...910...64K}, respectively. Panels e-h: the clusters identified by GMM, panels i-l: the clusters identified by FCM, and panels m-p: the clusters identified by BIRCH are plotted in the same order as the empirical IR-radio diagnostics. The clustering seems to be working well only in the case of KMeans (a-d), where we are able to compare the clusters and the empirical classifications. In the panels a-d, the purple cluster represents IR SFGs, the orange cluster represents IR AGN, and the green cluster represents radio AGN (see text for details). Each of these plots features two marginal histograms showing the normalised densities corresponding to the three clusters identified by each of the clustering tools.} 
    \label{fig:ir_radio_clusters}
\end{figure*}

GMM also requires the number of clusters to be predefined, but it is a soft clustering method, meaning every data point will be assigned a probability for it to belong to the respective cluster. Various fitting methods are available, but the expectation-maximisation (EM) algorithm \citep{dempster1977maximum} is widely used for finding the optimal clusters. In this work, we use the \verb|GaussianMixture| function from \verb|scikit-learn| for our analysis. One of the parameters that we adjusted was the \verb|covariance_type|. This parameter decides the covariance matrix of the Gaussian components, and hence the different shapes the clusters can adopt. We choose the \verb|covariance_type| of the \verb|GaussianMixture| function to be `diagonal' since this results in cluster distributions similar to other clustering tools. A choice of any of the available \verb|covariance_type|s results in similar cluster density distributions with only small rearrangements in the number of data points in the metallicity sequence (see \S\,\ref{sec:optical_clustering}).
\subsection{Fuzzy C-means}\label{sec:fcm}
Fuzzy sets as a class of objects with a continuum of membership grades was introduced by \cite{ZADEH1965338}. Since then, clustering problems based on fuzzy sets have been extensively studied \citep[see,][and references therein]{10.1007/978-81-322-2208-8_14}. Clustering algorithms that minimise an objective function and assign classes using the fuzzy technique are generally called fuzzy C-means \citep[FCM;][]{BEZDEK1984191,10.1007/978-81-322-2208-8_14}. FCM belongs to the set of partition-based methods, similar to KMeans in its clustering rationale, but in a fuzzy way, assigning probabilities (or the degree to which a given data point belongs to the clusters) to data points on their cluster assignments. FCM also require the number of clusters to be predefined. We use the \verb|Python| package \verb|fuzzy-c-means| \citep{dias2019fuzzy} in this work. The function has a parameter $m$ corresponding to the degree of fuzziness, which quantifies the degree to which each datapoint belongs to a given cluster component. For $m>2$, we found that the degree of fuzziness gets constrained between 0.3 and 0.7. This is not desirable since we want to explore the entire 0 to 1 range, as this range corresponds to a measure similar to a probability (although note that the degree of fuzziness parameter is not a probability). So we choose a value of $m=1.4$. This choice has implications when we calculate the probabilities in \S\,\ref{sec:prob_disc}.
\subsection{BIRCH}\label{sec:birch}
BIRCH \citep{Zhang1996BIRCHAE} is a hierarchical clustering algorithm built particularly for large datasets. Its main feature is that each clustering decision is made locally, that is, without scanning the entire dataset. This enables high speed clustering, especially beneficial for large datasets. It does so by using the characteristics of the clustering feature (CF) tree \citep[a compact representation of a cluster based on data statistics, see][for more details]{Zhang1996BIRCHAE}. The choice of the number of clusters and a parameter called \verb|threshold| plays a crucial role while using BIRCH. The parameter \verb|threshold| controls the resolution of each CF tree, where it acts as an upper limit for the CF tree components to split. These components are merged to form the desired number of clusters based on the predefined number of clusters. We use a value of 0.1 for \verb|threshold|, since lower values promote splitting of the clusters, and have resulted in cluster shapes comparable with astrophysical classifications (see \S\,\ref{sec:optical_clustering}). We use the \verb|BIRCH| function from \verb|scikit-learn| in \verb|Python| to perform clustering.
\section{Results}\label{sec:results}
The eight diagnostic diagrams described in \S\,\ref{sec:agn_classification} are investigated for clustering with the tools introduced in \S\,\ref{sec:clustering} and the results are presented in this section. In the discussions below, clusters refer to the groupings identified by the clustering tools and classes (or classifications) refer to the astrophysical classifications assigned by the AGN diagnostic tools. We tried various input parameter combinations for the clustering tools (different combinations of fluxes, magnitudes, colours, and flux ratios), and the optimal choice proved to be flux ratios and colours in the optical diagnostics and magnitudes (and radio fluxes) in the IR-radio diagnostics. %We have two options for the features to be fed into the clustering tools, either $(i)$ physical observables as they are (fluxes and magnitudes) or $(ii)$ combinations of the flux ratios and colours as used in the diagnostic diagrams. We find that the second option, flux ratios and colours, best facilitates the identification of clusters that we can easily associate with the SFG/AGN classifications. 
This is not surprising since these parameters represent the spaces where the data points exhibit segregation. We choose three as the number of clusters for better comparability with the astrophysical classifications. We find, using the functions associated with KMeans (the elbow method and the silhouette coefficient method), that the optimal number of clusters is either two or three, reinforcing our choice.

An optional step of dimensional reduction using PCA can be done before clustering. We investigated the effect of PCA on the clusters identified by different algorithms. This proved to be dependent on the respective spectral regime. The clusters identified in the optical diagnostic spaces compare well with the astrophysical classifications when the parameters do not undergo dimensional reduction. The clusters identified in the IR-radio spaces, on the other hand, compare well with the astrophysical classifications, if the principal components of the parameters are used. Essentially, optical-radio data do not benefit from PCA, whereas IR-radio data require PCA. It is not clear why this is the case. It could be because of the inherent differences in the optical and IR diagnostic parameter spaces, but exploration of this aspect is beyond the scope of the current work. PCA, when done, uses the \verb|Python| package \verb|pca| \citep{Taskesen_pca_A_Python_2020}, where the principal components carry at least 99\% of the variance. % As we show in the later parts of this section, whether or not this preliminary PCA step is helpful is subjective, and we state whether we employ PCA as an initial step where necessary.
As discussed in \S\,\ref{sec:data}, the optical-radio and the IR-radio data undergo clustering separately in their respective parameter spaces and these clusters are projected to the two-dimensional diagnostic spaces for analysis. 
\begin{figure*}[!t]
    \centering
    \includegraphics[width=\linewidth]{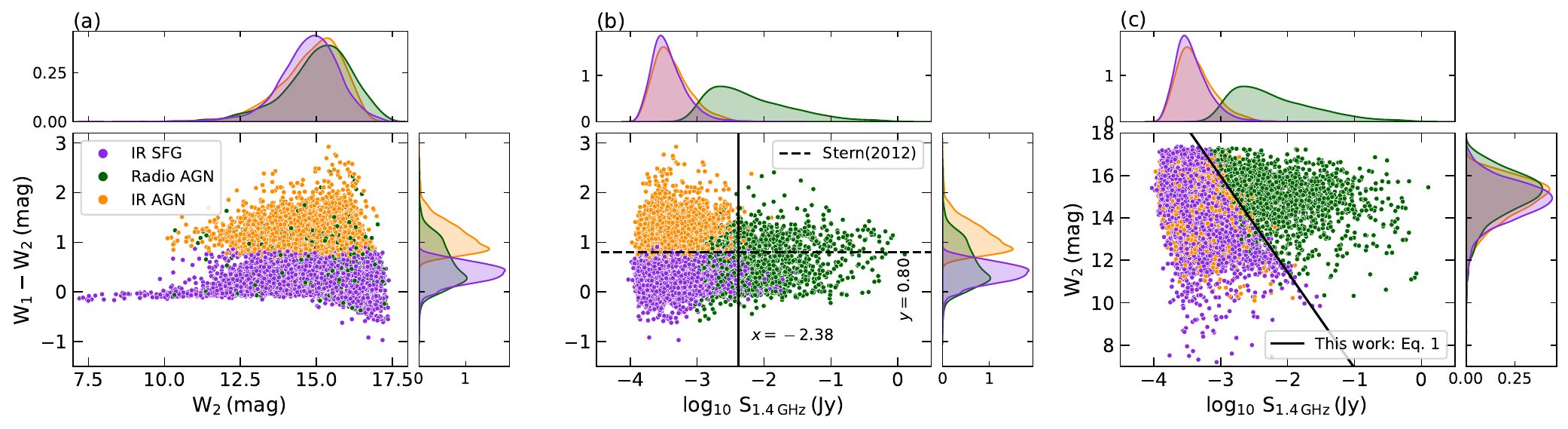}
    \caption{The distribution of KMeans clusters in various IR-radio spaces. Panel a shows the \cite{2018ApJS..234...23A} diagnostic space without the demarcation line, panel b shows the variation of the $\rm W_1-W_2$ colour as a function of \radioflux\,with the dashed line representing the \cite{2012ApJ...753...30S} demarcation between IR SFGs and IR AGN and the solid line at $\log_{10}S_{\rm1.4\,GHz}=-\,2.38\,\rm Jy$ separates the radio AGN from IR sources (see text for details). Panel c shows the $\rm W_2$ magnitude as a function of \radioflux, where the solid line (Equation\,\ref{eq:line}) separates the radio AGN from other sources. This plot is different from the \cite{2021ApJ...910...64K} diagnostic since they use the $\rm W_3$ flux. The colour scheme follows Figure\,\ref{fig:ir_radio_clusters}, but we are explicitly defining the purple cluster as IR SFGs, the orange cluster as IR AGN, and the green cluster as radio AGN, based on the characteristics evident from the discussions so far. The non normalised density of each of these clusters are shown as marginal densities following the same colour scheme.}
    \label{fig:2d_ir_radio}
\end{figure*}
\begin{figure}[!t]
    \centering
    \begin{subfigure}[b]{\textwidth}
    \centering
       \includegraphics[width=\linewidth]{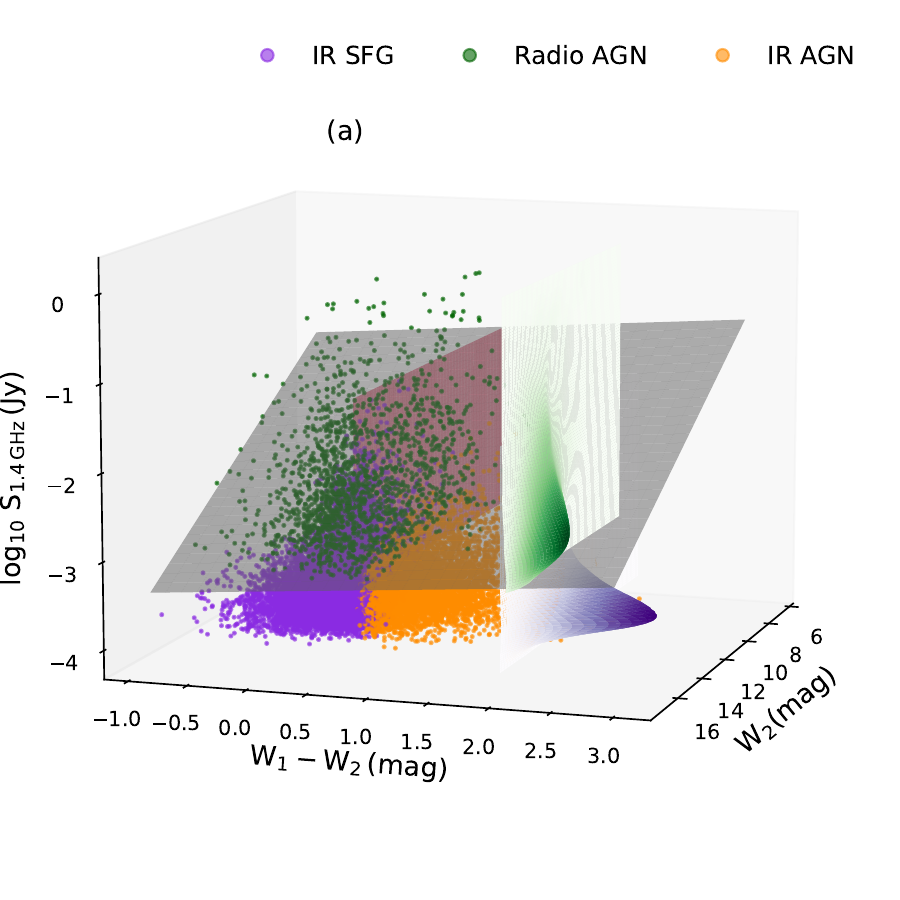}
    \end{subfigure}
    \begin{subfigure}[b]{\textwidth}
    \centering
        \includegraphics[width=\linewidth]{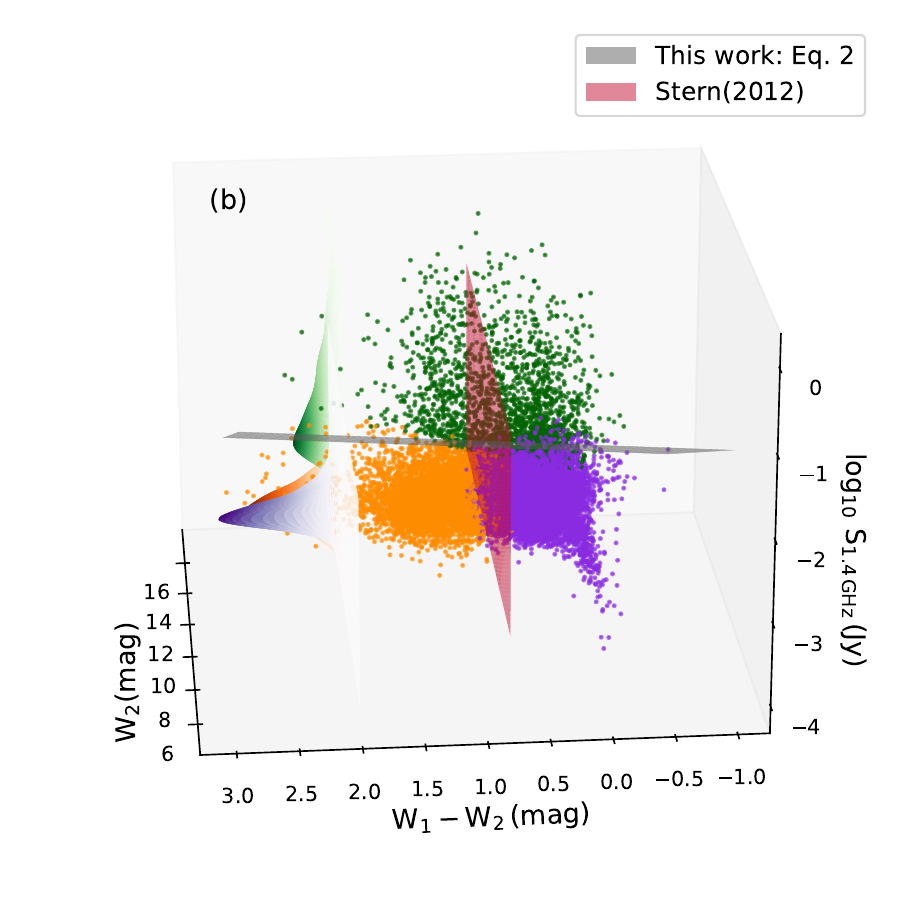}
    \end{subfigure}
    \caption{A three-dimensional IR-radio AGN diagnostic, combining the parameters shown in Figure\,\ref{fig:2d_ir_radio}. The colour scheme of the data points follows Figure\,\ref{fig:2d_ir_radio}, where the purple cluster represents the IR SFGs, the orange cluster represents the IR AGN, and the green cluster represents the radio AGN. The non-normalised densities of these species are shown following the same colour scheme. The grey plane, defined by Equation\,\ref{eq:3d_radio_plane}, separates the radio AGN from the IR sources with the resultant radio AGN being both complete and reliable over 90\% (see text for details). The crimson red plane corresponds to the three-dimensional version of the \cite{2012ApJ...753...30S} criterion separating the IR SFGs and IR AGN.}
    \label{fig:3d_ir_radio}
\end{figure}
\subsection{Optical-radio clusters}\label{sec:optical_clustering}
Figure \ref{fig:Optical_clusters} shows the clusters identified by the different algorithms in the optical-radio sample in a suite of optical diagnostics. The four panels show the clustering done by KMeans (\ref{fig:Optical_clusters}a-\ref{fig:Optical_clusters}d), GMM (\ref{fig:Optical_clusters}e-\ref{fig:Optical_clusters}h), FCM (\ref{fig:Optical_clusters}i-\ref{fig:Optical_clusters}l), and BIRCH (\ref{fig:Optical_clusters}m-\ref{fig:Optical_clusters}p), respectively. In every row there are four panels, each showing an optical diagnostic in the order BPT, MEx, blue, and CEx diagrams, respectively. For clustering these spaces, we choose six parameters: log(\logoiiihb), log(\logoiihb), log(\logniiha), stellar mass (log\,\logmstar), $u-g$ colour, and log\,\radioflux. The resulting clusters are then projected onto the two dimensional optical diagnostic spaces. These parameters are chosen as the combinations used in the relevant AGN diagnostics with the radio flux density as an additional constraint. %We choose the number of clusters to be three in each of these diagnostic tools, although there are standard methods to identify the optimum number of clusters (see \S\,\ref{sec:clustering}). This facilitates the comparison of the clusters with the astrophysical classes. 
Each clustering tool identifies three clusters distinguished by their colours: purple, green, and orange. The marginal plots show the density of each of these clusters for the corresponding axis parameter. No PCA was applied on the parameters for optical diagnostics since a dimensional reduction seemed less effective compared to the results presented here.

Upon visual inspection, it is evident that the three clusters identified by the different algorithms (along each row), as shown in Figure\,\ref{fig:Optical_clusters}, roughly follow the astrophysically motivated classifications (along each column). For instance, in Figure\,\ref{fig:Optical_clusters}a, we can see a clear distinction of the purple and green clusters into a star forming set towards smaller \logniiha\, and below the demarcation lines specified by \cite{2001ApJS..132...37K} and \cite{2003MNRAS.346.1055K}, and of the orange cluster into an AGN set above the \cite{2001ApJS..132...37K} line. The clusters that likely represent the SFG population is divided into two (purple and green dots), which most likely correspond to a metallicity sequence where the purple cluster represents low-metallicity star formers and the green cluster represents the high-metallicity star formers. This is supported by the separation of these clusters as a function of galaxy stellar mass \citep[the mass-metallicity relation;][]{2004ApJ...613..898T}, in the MEx diagram shown in panels \ref{fig:Optical_clusters}b, \ref{fig:Optical_clusters}f, \ref{fig:Optical_clusters}j, and \ref{fig:Optical_clusters}n. This split of the SFG population into a metallicity sequence is most likely driven by the data distribution in the six dimensional parameter space used for clustering.

Even though clusters are distinguishable when compared to the demarcation lines, they are not completely separate, as evident from the marginal density distributions. We do not expect both statistical clusters and astrophysical classifications to be identical since both techniques treat the parameter space uniquely. The overlap between the multiple clusters, shown consistently throughout the clustering tools, might be an indication of a softer boundary between the populations, a concept we explore throughout in this work.
\subsection{Infrared-radio clusters}\label{sec:ir_clustering}
Figure \ref{fig:ir_radio_clusters} shows three IR AGN diagnostics, in the first three columns, and the radio AGN diagnostic in the last column. It shows the action of KMeans (\ref{fig:ir_radio_clusters}a - \ref{fig:ir_radio_clusters}d), GMM (\ref{fig:ir_radio_clusters}e - \ref{fig:ir_radio_clusters}h), FCM (\ref{fig:ir_radio_clusters}i - \ref{fig:ir_radio_clusters}l), and BIRCH (\ref{fig:ir_radio_clusters}m - \ref{fig:ir_radio_clusters}p), respectively, on these diagnostic spaces. Clustering is done on a four-dimensional IR-radio space containing the three WISE magnitudes, $\rm W_1,\ W_2,\ W_3$ and log\,$S_{1.4\,{\rm GHz}}$. We do not use K-band magnitude as a clustering parameter. As described in \S\,\ref{sec:final_sample}, only around 30\% of the IR-radio sample has K-band magnitude available. Thus, including the K-band magnitude as a parameter may bias the clustering results. The clustering rationale and the figure characteristics are similar to \S\,\ref{sec:optical_clustering} where the three clusters are distinguished by their colours: purple, green, and orange. The clustering in this case, however, is done on the principal components carrying 99\% variance of the astrophysical features, as they showed clearer separation.

It is evident from Figure\,\ref{fig:ir_radio_clusters} that the KMeans clusters are well distinguishable and populate the diagnostic space similar to the astrophysical classifications in the IR-radio diagnostics. The purple and orange clusters occupy the IR SFG and AGN regions, respectively, in the IR diagnostics (Figures \ref{fig:ir_radio_clusters}a-\ref{fig:ir_radio_clusters}c). Notably, we see that the orange clusters occupy predominantly the same regions (above the \citealt{2018ApJS..234...23A} line, within the \citealt{2012MNRAS.426.3271M} wedge, and the \citealt{2012ApJ...754..120M} wedge in columns a, b, and c, respectively) as those of the AGN classes from the diagnostics. The IR diagnostics considered here do not have a composite classification, and the green cluster is dispersed through the former two clusters. It is interesting to see that this dispersed population, however, occupies the region identified as AGN in the radio diagnostic in Figure\,\ref{fig:ir_radio_clusters}d. This feature is quite important, with the potential to explore radio AGN those are not flagged as IR AGN. We explore this in more detail below in \S\,\ref{sec:2d_ir_radio}. Both purple and orange clusters (SFGs and AGN, respectively, in the IR diagnostics) occupy the SFG region of the radio diagnostic in Figure\,\ref{fig:ir_radio_clusters}d. This means that the orange cluster represents a radio-quiet IR active galaxy population.
It is clear that GMM, FCM, and BIRCH perform poorly in identifying clusters that can be compared with the astrophysical classes. This is evident from the Figures \ref{fig:ir_radio_clusters}e-\ref{fig:ir_radio_clusters}p since the different clusters occupy regions that are not differentiated by any demarcation lines of the relevant IR-radio diagnostics.

Both the first and third columns in Figure~\ref{fig:ir_radio_clusters}, corresponding to the \cite{2018ApJS..234...23A} and \cite{2012ApJ...754..120M} diagnostics, feature an extended tail towards $\rm W_2\lessapprox12$ and $\rm W_1-W_2\approx0.2$. These regions in these parameter space are predominantly occupied by star-like objects, and a subsample was visually inspected to assess the stellar proportion. There are only 303 (out of 28\,462, 1\%) objects satisfying these criteria. For visual inspection, we crossmatched these 303 objects with the GAMA database at a 5" radius, resulting in 202 GAMA objects. For the remaining sources, we used the archival data from the Barbara A. Mikulski Archive for Space Telescopes (MAST)\footnote{\hyperlink{mast.stsci.edu}{mast.stsci.edu}}. It was found that there are star-like objects in our sample in this region. 
However, many objects here are also found to be extended sources. Excluding stellar objects using a selection criterion like a lower bound on redshift is not possible in this case since we do not have redshift information available for many sources in the IR-radio sample. Using a magnitude cut-off would also result in the removal of true galaxies. As we are more interested in the green cluster (see next section) and there are only 16 of these in the star-like region (of which none appear point-like), we do not remove these objects from the analysis. The small numbers involved ensure their inclusion does not significantly affect our results or conclusions.

\subsection{An IR-radio AGN diagnostic}\label{sec:2d_ir_radio}
Out of the three populations identified by KMeans in Figure\,\ref{fig:ir_radio_clusters}a, the purple and orange clusters (representing IR SFGs and AGN) lie below and above the demarcation line, respectively. The green cluster is distributed broadly across the parameter space. This scatter is not clear from the marginal density plots because the green cluster members correspond to only 6\% of the total population, and the density plots are normalised with respect to the total number of sources. It was noted in \S\,\ref{sec:ir_clustering} that the green cluster that appears dispersed in the IR diagnostic space (Figure\,\ref{fig:ir_radio_clusters}a) populates the AGN space in the radio diagnostic, as shown in Figure \ref{fig:ir_radio_clusters}d. To investigate if there exists any natural grouping between these clusters, we analyse a combination of the IR-radio parameter spaces using the KMeans clusters, shown in Figure\,\ref{fig:2d_ir_radio}. The colours of the clusters follow those in Figure\,\ref{fig:ir_radio_clusters}a-\ref{fig:ir_radio_clusters}d, where purple clusters represent IR SFGs, orange clusters represent IR AGN, and green clusters represent radio AGN, all identified using KMeans as described in \S\,\ref{sec:ir_clustering}. The marginal density plots are not normalised with respect to each other to better aid the visual distinction of the clusters. 

Figure\,\ref{fig:2d_ir_radio}a reproduces the \cite{2018ApJS..234...23A} diagnostic (as shown in Figure\,\ref{fig:ir_radio_clusters}a). Figure\,\ref{fig:2d_ir_radio}b shows the $\rm W_1-W_2$ colour as a function of $S_{1.4\,{\rm GHz}}$ with the dashed line \citep{2012ApJ...753...30S} separating the IR SFGs from the IR AGN. Above the solid black line at $\log_{10}S_{\rm1.4\,GHz}=-\,2.38\,\rm Jy$ sources are identified as radio AGN (green dots) with only 10\% contamination from IR detected sources. It is well-known that for flux-densities brighter than a few mJy at this frequency the radio source population is dominated by AGN \citep[e.g.,][]{2008MNRAS.386.1695S}.

Figure\,\ref{fig:2d_ir_radio}c is a variant of the diagnostic defined by \cite{2021ApJ...910...64K}, also shown in Figure\,\ref{fig:ir_radio_clusters}d, and is of special interest to us. Instead of the $\rm W_3$ flux, it looks at the variation of the $\rm W_2$ magnitude (since we use the $\rm W_1-W_2$ colour and we can benefit from a sensitivity better than the $\rm W_3$ magnitude)  as a function of \radioflux. Notably, we see that the combination of these two parameters is able to separate the radio population from the IR-detected sources.
%Unlike Figure\,\ref{fig:2d_ir_radio}b, the IR AGN and SFGs cannot be distinguished in Figure\,\ref{fig:2d_ir_radio}c, although, an IR-radio separation is clear in this parameter space.
The solid black line, defined by
\begin{equation}
    {\rm W}_2 = m\,\log_{10}\,S_{1.4\,{\rm GHz}}+c,
    \label{eq:line}
\end{equation}
where the slope, $m=-\,4.519$ and the y-intercept, $c=2.409$, separates the IR and radio population, such that the radio AGN lying above the line defined in Equation\,\ref{eq:line} is 90.26\% reliable and 91.25\% complete. It is important to note the caveats here. Being a flux-cut, it is likely that the numbers stated are specific to this sample and survey. A more sensitive radio survey can, for the same threshold, identify sources with varying reliability and completeness. Also, the completeness and reliability values reported here are based only on radio AGN. These values will increase slightly if we account for both IR and radio AGN.
\subsection{A three-dimensional perspective}\label{sec:3d_ir_radio}
The previous section introduced a new IR-radio AGN selection, Equation\,\ref{eq:line}, which is relatively reliable and complete. %In terms of the diagnostic spaces, Figure\,\ref{fig:2d_ir_radio}c does not carry the IR-radio separation exhibited in Figure\,\ref{fig:2d_ir_radio}b.
In this section, we build on this two-dimensional diagnostic with a three-dimensional approach.

Figure\,\ref{fig:3d_ir_radio} shows a three-dimensional space composed of the \cite{2018ApJS..234...23A} (Figure \ref{fig:ir_radio_clusters}a) parameters as the x- and y-axes, and $\log_{10}$\,\radioflux\,as the z-axis. Figures \ref{fig:3d_ir_radio}a and \ref{fig:3d_ir_radio}b show the different orientations of the same parameter space. The data points represent the KMeans clusters identified in the Figures \ref{fig:ir_radio_clusters}a-\ref{fig:ir_radio_clusters}d, where the purple clusters represent IR SFGs, orange clusters represent IR AGN, and the green clusters represent radio AGN. The (non-normalised) densities of each of these clusters are shown on the white surfaces with projections mapped to the same colours as the data points. A clear separation of the different populations is apparent in these diagrams.

We define planes of demarcation between the different clusters in Figure\,\ref{fig:3d_ir_radio}. The grey plane separates the green clusters (corresponding to radio AGN) from the others (representing IR galaxies) and is defined as 
\begin{equation}
    a\,{\rm W}_2+b\,({\rm W}_1-{\rm W}_2)+c\,S_{1.4\,{\rm GHz}}+d=0,
    \label{eq:3d_radio_plane}
\end{equation}
where the coefficients, $a=0.4598,\,b=-\,0.2491,\,c=2.3541$, and $d=-\,0.2201$ are chosen such that the contamination by the clusters representing the IR sources above this plane is constrained to a maximum of $\approx$\,10\%. This criterion selects radio AGN such that in the region where Equation\,\ref{eq:3d_radio_plane} > 0, the selection is 92.22\% reliable and 91.49\% complete. A second criterion that separates the clusters representing the IR SFGs and IR AGN is shown as the crimson plane, which is simply a two-dimensional version of the \cite{2012ApJ...753...30S} criterion.
\begin{figure*}
    \centering
    \includegraphics[width=0.9\linewidth]{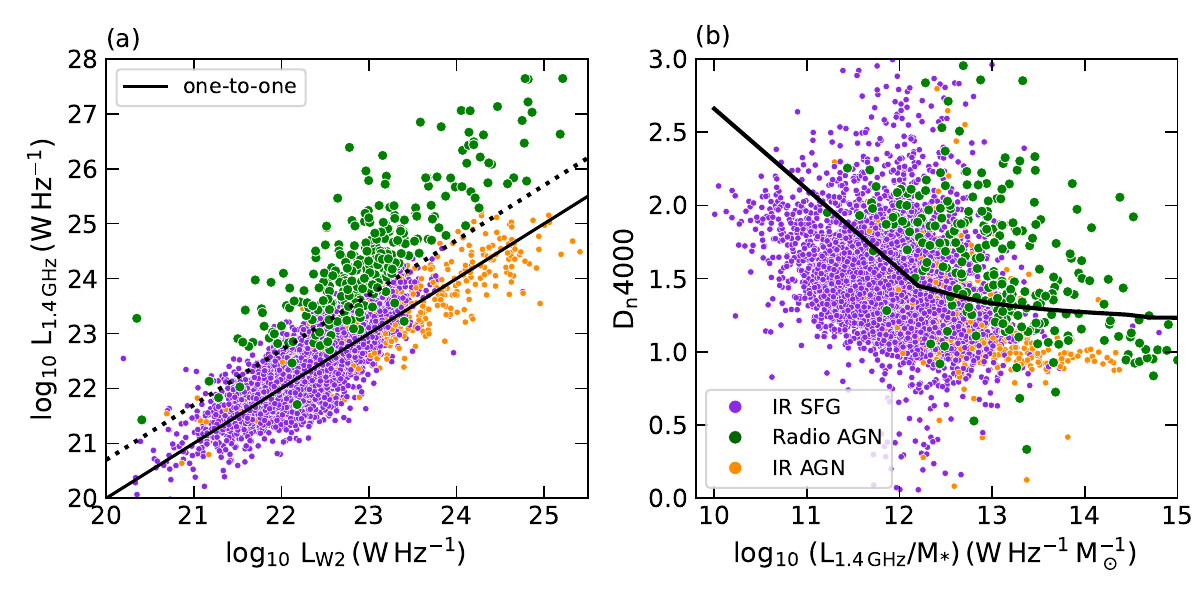}
    \caption{The figure shows the distributions of the IR-radio sources with spectroscopic detections (contains 32\% of the IR-radio sample). Panel (a): The IR-radio correlation between the $\rm W_2$ magnitude and the 1.4 GHz radio flux density. The colours of the dots follow the previous figures. The solid line is the linear relation between the two quantities, and the dotted line corresponds to a version scaled by a factor of 5 (in linear scale). Around 88\% of the radio AGN (green dots) lie above the dotted line, implying that the radio emission from a large fraction of the sample is powered by AGN activity. Panel (b): The $\rm D_{n}4000$ break index as a function of the 1.4 GHz radio flux density per stellar mass, as used by \cite{2012MNRAS.421.1569B}. The objects falling below the solid line are most likely low-excitation in nature. It is evident that the radio AGN are scattered around this demarcation line, implying that the optically detected IR-radio sources are of both low- and high-excitation nature.}
    \label{fig:firrc}
\end{figure*}
The high completeness and reliability of the radio AGN selection in the three-dimensional criterion, while maintaining the IR SFG-AGN separation, demonstrates the significant potential of higher-dimensional diagnostics for AGN classification.

The parameter space in Figure\,\ref{fig:3d_ir_radio} is essentially a combination of the best features of Figures\,\ref{fig:2d_ir_radio}b and \ref{fig:2d_ir_radio}c. There are multiple benefits to such a selection criterion (Equations\,\ref{eq:line} and \ref{eq:3d_radio_plane}), $(i)$ it encodes information from two spectral regimes, namely, IR and radio, a benefit of the clustering approach; $(ii)$ since these criteria do not need optical counterparts for comparison, and the IR counterparts of the radio sources are more commonly identified, the sample for AGN identification is no longer constrained by its size, thereby increasing the number of available candidates; $(iii)$ the separation of IR sources into SFGs and AGN is still evident; $(iv)$ a stricter radio selection resulting in radio sources where the nuclear black hole activity most likely contributes to a larger fraction of the radio emission; and $(v)$ enables the selection of sources that are likely active in one spectral regime and not the other, for instance, radio AGN those would otherwise be identified as IR star formers. An implication of the last point is that the diagnostic has the potential for selecting low excitation radio galaxies \citep[LERGs;][]{2014ARA&A..52..589H,2024PASA...41...16P}, a class of AGN that are active primarily in the radio. 

Figures\,\ref{fig:firrc}a and \ref{fig:firrc}b test the last two points. They use the optical spectra and redshifts of the IR-radio sample (where available) to analyse IR-radio correlation in Figure\,\ref{fig:firrc}a, and the low-excitation nature of these galaxies, as prescribed by \cite{2012MNRAS.421.1569B} in Figure\,\ref{fig:firrc}b. This sample of 9\,027 sources (32\% of the IR-radio sample) is constructed by cross-matching the IR-radio sample with the GAMA spectroscopic data with a 5" radius. The dotted line in Figure\,\ref{fig:firrc}a is 0.7 dex (5 in linear scale) above the one-to-one (solid) line \citep[see e.g.][for a similar selection of radio-excess AGN, but with a lower threshold value of three in the linear scale]{2017A&A...602A...3D,2024PASA...41...16P}. This is a conservative threshold so that the radio AGN selection minimises contamination. Around 88\% of the green dots (radio AGN) lie above this line, supporting the point that AGN activity contributes to a large fraction of the radio emission. It is also worth noting that the IR-radio correlation evolves with redshift \citep[see][]{2016MNRAS.457..629C}. This evolution would not affect our results since the effect is more prominent for $z>1$ \citep{2021A&A...647A.123D}. LERGs are expected to lie below the solid line in Figure\,\ref{fig:firrc}b (see \citealt{2012MNRAS.421.1569B} for more details). It is clear that a large part of the radio AGN population lies above this line. This implies that the sources detected in optical, IR, and radio regimes are not predominantly LERGs. This is expected, as the requirement of optical spectra biases against optically weak sources, potentially removing LERG-like objects. A consensus on the LERG population seems not feasible without more sensitive spectroscopic measurements.

Mid-IR AGN selection techniques, especially comprehensive approaches such as those of \cite{2012ApJ...753...30S} and \cite{2018ApJS..234...23A}, are highly effective at identifying clean samples of luminous AGN. However, they are biased against low-luminosity sources, particularly at higher redshifts \citep{2012ApJ...748..142D,2012MNRAS.426.3271M,2014MNRAS.438..494R}. Consequently, their results cannot be straightforwardly generalised to the full AGN population. Here, our novel approach stands out. A selection, combining mid-IR and radio observations like ours, can be used to partly rectify this bias, but deeper observations are needed for a better understanding.
%A comparison with our optically selected radio sample reveals that the radio AGN selected by the  \cite{2021ApJ...910...64K} diagnostic represent only 3\% of the total optical-radio sample, whereas the cluster corresponding to radio AGN in the new IR-radio diagnostic (Figure\,\ref{fig:3d_ir_radio}) represent 6\% of the total IR-radio sample. This may look like a small number when compared to the 18\% radio AGN detected by the \cite{2021ApJ...910...64K} diagnostic until we think about the possible contamination by IR star formers and AGN. As already stated, the new selection is potentially a stricter criterion that discards most likely IR dominant sources, resulting in a much robust population of radio AGN with better numbers when drawn upon data probing deeper Universe with lesser selection constraints. %A potential bias here is the low excitation radio galaxies \citep[LERGs;][]{2014ARA&A..52..589H,2024PASA...41...16P} where it may not be possible to establish a correlation between the emissions at the radio and other wavelengths. Although, this depends on the exact definition adopted for LERGs (whether they are inactive only in the optical or both in the optical and IR). 
\section{Discussion}\label{sec:discussion}
\subsection{Clusters and classifications}\label{sec:cluster_disc}
\subsubsection{Optical parameter space}\label{sec:optical_disc}
\begin{table*}[!ht]
    \centering
    \begin{tabular}{ccc|cc|cc|cc|cc}%SSSSSSSSSSS
    \toprule
    \multicolumn{3}{c}{Optical} &
      \multicolumn{2}{c}{KMeans (\%)} &
      \multicolumn{2}{c}{GMM (\%)} &
      \multicolumn{2}{c}{FCM (\%)} &
      \multicolumn{2}{c}{BIRCH (\%)} \\
      & & & SFG & AGN & SFG & AGN & SFG & AGN & SFG & AGN \\
      \midrule
      BPT & SFG & 3\,467 & 94.89 & 5.11 & 91.90 & 8.10 & 87.68 & 12.32 & 93.37& 6.63\\
          & AGN & 817  & 15.91 & 84.09 & 21.30 & 78.70 & 7.59 & 92.41 & 10.41 & 89.59\\\hline
      {MEx} & SFG & 2\,514 & 98.58 & 1.42 & 93.34 & 6.66 & 97.15 & 2.85 & 97.11 & 2.89\\
          & AGN & 2\,106 & 60.45 & 39.55 & 63.01  & 36.99 & 46.82 & 53.18 & 57.31 & 42.69\\\hline
      {Blue} & SFG & 4\,547 & 88.15 & 11.85 & 85.92 & 14.08 & 79.85 & 20.15 & 85.92 & 14.08\\
           & AGN & 512 & 18.56 & 81.44 & 23.24 & 76.76 & 17.58 & 82.42 & 15.82 & 84.18\\\hline
      {CEx} & SFG & 3\,346 & 96.08 & 3.92 & 91.87 & 8.13 & 89.51 & 10.49 & 93.84 & 6.16\\
            & AGN & 938 & 21.86 & 78.14 & 30.49 & 69.51 & 11.41 & 88.59 & 19.4 & 80.60\\
      \bottomrule
    \end{tabular}
    \caption{The distribution of the SFGs and AGN identified by the different optical diagnostic tools among the various clustering tools, with each cell acting as a confusion matrix for a diagnostic-clustering tool pair. The two clusters in the metallicity sequence exhibited by the SFG population (purple and green clusters) have been combined to form a single SFG cluster in the case of the clustering tools. The diagnostic tools are arranged along the rows in the order BPT, MEx, blue, and CEx, and the clustering tools are arranged along the columns in the order KMeans, GMM, FCM, and BIRCH. Approximately 90\% of the star forming population identified by the empirical tools are correctly identified by most clustering tools. This fraction comes down to 80\% in the case of AGN, with even lower values for the MEx diagnostic (second row, see text for details).}
    \label{tab:comparison}
\end{table*}
Table\,\ref{tab:comparison} evaluates the performance of the various clustering tools on different optical AGN diagnostics. It is clear from the discussions so far that the purple and green clusters from the clustering tools represent the SFG class and the orange clusters represent the AGN class from the astrophysical classifications. Accordingly, we label the purple and green clusters as SFGs and the orange clusters as AGN in the discussions in the current and following sections. In Table\,\ref{tab:comparison}, the composite class from the diagnostic tools and the metallicity sequence identified by the clustering tools (purple + green) are labelled as their respective SFG classes. That is, the SFG label under the diagnostics (along the rows) include both SFGs and composites and the SFG label under the clustering tools (along the columns) contain both the non AGN clusters.

In the case of SFGs, along the rows in Table\,\ref{tab:comparison}, it is evident that in most cases $\approx$\,90\% of sources are correctly identified by the clustering tools. Clusters and classes among AGN match at best for $\approx$\,90\% of the sources, with the number dropping to $\approx$\,40\% for the MEx diagnostic (second row of Table\,\ref{tab:comparison}). To investigate whether this mismatch stems from using \radioflux\,as one of the features, we checked the fractions shown in Table\,\ref{tab:comparison} in a parameter space comprised only of optical parameters. No notable difference was detected in the numbers except for small rearrangements in the fractions. This insensitivity to the radio flux density could be interpreted as the mutual exclusion between the astrophysical origins of the optical and radio AGN activity for the sources in our sample. This is no surprise since optical source selection discards a large fraction of the absorption line galaxies, which primarily represent the radio galaxy population, at least in the local Universe \citep[say $z<0.4$, e.g.][]{2017A&A...602A...2S,2023MNRAS.523.5292K}. This is evident from the number of radio sources with and without optical counterparts used for clustering. From Table\,\ref{tab:final_sample} and \S\,\ref{sec:emu_data}, it can be seen that only $\approx$\,5\% of the total radio sources in both G09 and G23 fields present spectral lines for performing optical diagnostic analysis. This can only be improved using a large scale, deeper spectroscopic survey in conjunction with the EMU data.
\begin{table}[!htb]
    \centering
    \begin{tabular}{ccc|cc}%SSSSSSSSSSS
    \toprule
    \multicolumn{3}{c}{Infrared-Radio} &
      \multicolumn{2}{c}{KMeans (\%)} \\
      & & & SFG & AGN \\
    \midrule
    Assef(2018) & SFG & 22\,400 & 93.33 & 6.67 \\
                & AGN & 4\,152 & 14.38 & 85.62 \\
    \hline
    Mateos(2012)& SFG & 22\,767 & 94.28 & 5.72 \\
                & AGN & 3\,785 & 0.98 & 99.02 \\
    \hline
    Messias(2012)& SFG & 25\,876 & 81.71 & 18.29 \\
                 & AGN & 676 & 52.96 & 47.04 \\
    \hline
    Kozie{\l}-Wierzbowska(2021)& SFG & 22\,925 & 99.98 & 0.02\\
                & AGN & 5\,537 & 47.51 & 52.49\\
      \bottomrule
    \end{tabular}
    \caption{The distribution of the SFGs and AGN identified by the different IR-radio diagnostic tools among the KMeans clusters, with each cell acting as a confusion matrix for a diagnostic-clustering tool pair. It is worth noting the nearly 0\% misclassification of the \cite{2021ApJ...910...64K} radio SFGs as KMeans AGN. The \cite{2021ApJ...910...64K} radio AGN has a success rate of $\approx$\,52\%, which makes the KMeans clustering selection a stricter, less contaminated AGN selection. The numbers noted in the first three rows (the IR AGN diagnostics) do not include the green cluster population (see text for details).}
    \label{tab:ir_comparison}
\end{table}
\subsubsection{IR and radio parameter spaces}\label{sec:ir_radio_disc}
Table\,\ref{tab:ir_comparison} looks at the distribution of sources among the different IR and radio empirical diagnostic tools (see \S\,\ref{sec:ir_agn} and \S\,\ref{sec:radio_agn}) and the KMeans clustering tool. We do not include other clustering tools here since they do not seem to perform as efficiently as KMeans. We attribute this differing behaviour of the clustering tools to the distribution of the astrophysical classes in the various diagnostic diagrams. The classification schemes in the optical diagnostic diagrams are mostly driven by the density distribution of the data points. Since most clustering tools search for such density distributions, the resulting clusters, for the most part, would agree with the classifications. IR demarcations, on the other hand, are specific cut-offs, which is likely why clustering tools other than KMeans struggle in identifying the astrophysical classification. KMeans does not search for density distributions; rather, it can identify threshold cut-offs as seen in Figure\,\ref{fig:ir_radio_clusters}a. It should be noted here that there could be other clustering tools those may have worked as efficiently as KMeans in the scenario presented in this work.

It is evident by now that, in the case of the IR diagnostics, the purple and orange clusters represent the IR SFGs and AGN, respectively. The green clusters, on the other hand, represent the radio AGN as seen in the radio diagnostic (Figure\,\ref{fig:2d_ir_radio}). The numbers reported in Table\,\ref{tab:ir_comparison} take this into account. The totals in the first three rows do not include the green cluster identified in the respective diagnostics (Figures\,\ref{fig:ir_radio_clusters}a-\ref{fig:ir_radio_clusters}c). The last row of Table\,\ref{tab:ir_comparison} includes all three clusters identified in the radio diagnostic (Figure\,\ref{fig:ir_radio_clusters}d). This is because in the clustering results, there is no KMeans label for the green data points that can be used to put them into either IR SFG or AGN classes. Along the rows, SFGs are identified correctly in at least 90\% of the cases, except for the \cite{2012ApJ...754..120M} criterion, where the identification rate is $\approx$\,80\%. This can be attributed to the requirement of K-band magnitudes resulting in a reduced sample size for this diagnostic (see \S\,\ref{sec:final_sample}). Also, K-band magnitude was not a clustering parameter, which places a weaker constraint on the cluster identification. These two reasons might be playing the role in driving the difference we see, given that this is a statistical study. AGN classes show varying matching rates, mainly above 80\% for the \cite{2018ApJS..234...23A} and \cite{2012MNRAS.426.3271M} criteria. For the \cite{2012ApJ...754..120M} diagnostic, where the matching rate reaches $\approx$ 47\%, a similar explanation as for the SFGs applies here.
\begin{figure}[!t]
    \centering
    \includegraphics[width=\linewidth]{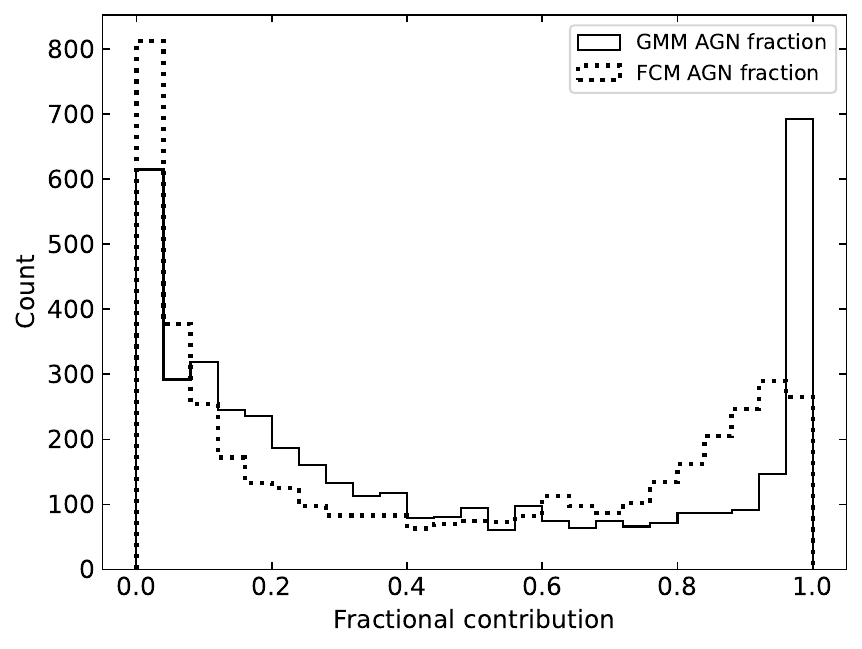}
    \caption{Fractional contributions (or probabilities) of AGN activity and the corresponding number of sources, as reported by GMM and FCM. The solid and dashed lines correspond to the AGN fractions from GMM and FCM, respectively. Since a two cluster approach is adopted, AGN and SFG probabilities in a given clustering tool add up to one. These values can be compared to the relative strengths of the energy generating mechanisms in a given galaxy. As a result,  instead of labelling a galaxy either as an SFG or AGN, it can be given a fractional quantification. That is, for instance, 40\% SFG and 60\% AGN. In a hard classification setting, such a galaxy would most probably be classified as an AGN, still carrying the unaccounted contribution from star formation processes. Methods similar to this work can be used to soft classify galaxies such that the contributions are accounted for without any bias.}
    \label{fig:opt_probdist}
\end{figure}
\begin{figure*}[!ht]
    \centering
    \includegraphics[width=0.8\linewidth]{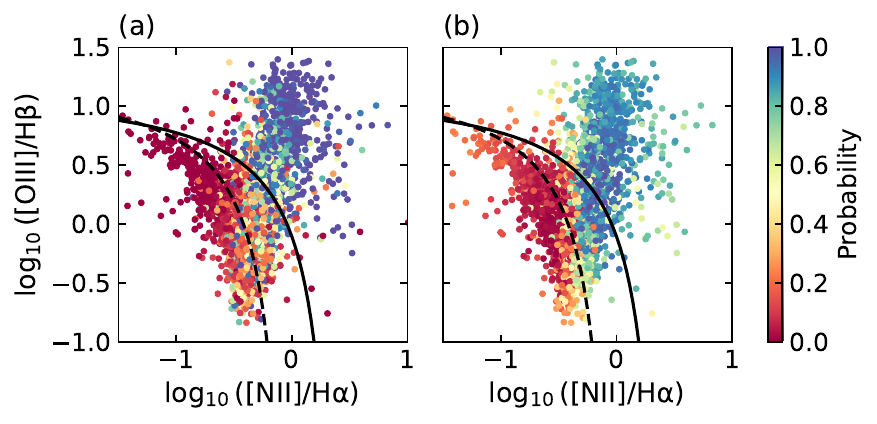}
    \caption{The distribution of the probabilities described in Figure\,\ref{fig:opt_probdist} in the BPT parameter space, in the order, GMM AGN in panel a and FCM AGN in panel b. It can be seen that the AGN exhibit higher probabilities in their respective BPT space. The solid and dashed lines correspond to the \cite{2001ApJS..132...37K} and \cite{2003MNRAS.346.1055K} lines, respectively.}
    \label{fig:opt_prob_bpt}
\end{figure*}

It is worth noting the distribution of empirical radio AGN among the KMeans AGN cluster (last row), where the matching rate is close to 52\%, which appears relatively low. This, in fact, corresponds to a much stricter radio AGN selection criterion than a simple one-to-one relation prescribed by \cite{2021ApJ...910...64K} (also see \S\,\ref{sec:ir_clustering}), which is made clear by the negligible (close to 0\%) fraction of \cite{2021ApJ...910...64K} SFGs labelled as KMeans AGN. The remaining $\approx$\,47\% of sources belong to radio star formers (orange and purple clusters) in the KMeans classification, out of which $\approx$\,56\% are IR AGN and $\approx$\,37\% are IR SFGs. A less conservative criterion would redistribute these sources among radio AGN, thereby raising the KMeans radio AGN percentage with respect to \cite{2021ApJ...910...64K} AGN. This point implies that we can have high confidence in the radio AGN population selected by the IR-radio AGN diagnostics defined in \S\,\ref{sec:2d_ir_radio} and \S\,\ref{sec:3d_ir_radio}, since they are defined to select KMeans clusters corresponding to radio AGN.
\subsection{A probabilistic view on emission mechanisms}\label{sec:prob_disc}
\subsubsection{Optical AGN probabilities}\label{sec:optical_probs}
Traditional galaxy classification methods work by identifying the dominant emission mechanism, resulting in binary labels for all the galaxies in the sample, a procedure followed throughout this work so far. The simplest, and yet the key point here, is that the contributions from the non-dominant component are still present in the galaxy, and are overlooked. Recent explorations in this direction are provided by \cite{2014MNRAS.444.3961D,2014MNRAS.439.3835D,2018ApJ...861L...2T}, and \cite{2025MNRAS.537.1028P}. \cite{2018ApJ...861L...2T} explored the degree of mixing in the optical spectra of SFGs and AGN using photoionisation models. They found that a median of $\approx$\,30\% of the Balmer line flux arises in star forming like regions, even for galaxies with high values of excitation ($\rm\log$\,\logoiiihb$\geq0.9$). \cite{2025MNRAS.536L..32M} recently used brightness temperature thresholds to put an upper limit on the radio emission from star formation, thereby isolating regions of radio emission from AGN to derive 144\,MHz radio luminosity functions.

In this work, we follow a similar philosophy, where a probabilistic (or fractional) quantification of the different components is preferred over a binary classification of the galaxy energy budget. Some of the clustering tools, such as GMM and FCM, can produce probabilities associated with astrophysical classifications. We infer these probabilities as related to the existence of AGN. As already described in \S\,\ref{sec:gmm} and \S\,\ref{sec:fcm}, GMM considers the data points in the feature space as Gaussian probabilities and assigns the most probable cluster label as a hard classification. FCM follows a similar approach, however, it gives the degree to which a data point belongs in a given cluster (in a fuzzy way, see \S\,\ref{sec:fcm}) and assigns the cluster label with the highest degree to that data point. The main difference between these two algorithms is that GMM assumes cluster Gaussianity, whereas FCM relies solely on cluster centroid distances. We choose to keep FCM as well in this discussion since the clusters may not always be Gaussian in reality.

We argued in \S\,\ref{sec:optical_clustering} that the two clusters lying below the \cite{2001ApJS..132...37K} line (purple and green) represent star forming species exhibiting the metallicity sequence. We combine the probabilities of these two clusters by performing the clustering for the optical sources using the number of clusters to be two, and present the fractional contribution of star formation and nuclear activity. Two probability values are assigned to each data point by both GMM and FCM, corresponding to their occupancy in one of the two clusters. Figure\,\ref{fig:opt_probdist} shows the AGN fractions identified by GMM (solid line) and FCM (dashed line). Both clustering tools recover similar fractions of AGN and SFGs in the optical-radio sample. There is, however, a difference in the number of galaxies identified by both algorithms when it comes to pure SFGs and AGN. This can be attributed to the working principles of GMM and FCM. GMM models the data density, whereas FCM looks at the distances between the cluster centroids and data points. When the clusters strongly overlap, GMM takes the data points in that region to be genuinely belonging to both the clusters, assigning similar probabilities. FCM, on the other hand, looks for the centroid closest to that data point and assigns more weight to that cluster.

Figure\,\ref{fig:opt_prob_bpt} shows the distribution of these probabilities in the BPT space, where Figures\,\ref{fig:opt_prob_bpt}a and \ref{fig:opt_prob_bpt}b correspond to the GMM and FCM AGN probabilities, respectively. The colour is mapped to the probability for a given data point to belong to the cluster representing AGN. Since we followed a two-cluster approach, the total probability of AGN and SFGs from a given clustering tool adds up to one. It is evident that AGN exhibit higher probabilities in their respective regions in the BPT diagram. Both GMM and FCM probabilities exhibit strong separation between the SFG and AGN regions, such that the sources with higher excitation values present the maximum probability to be AGN. The probabilities assigned by the clustering techniques look remarkably like a model for the AGN-SFG mixing, similar to the mixing line defined by \cite{2018ApJ...861L...2T}, where the data points near the dividing lines have a higher probability to be SFGs and the data points with high excitation values show higher probabilities to be AGN. The similarity between the probability measures from the clustering and astrophysical modelling strengthens the results from clustering, such that these measures can be directly associated with the strengths of the underlying galaxy energy sources.
\subsubsection{IR AGN probabilities}\label{sec:ir_probs}
\begin{figure}[!htb]
    \centering
    \includegraphics[width=\linewidth]{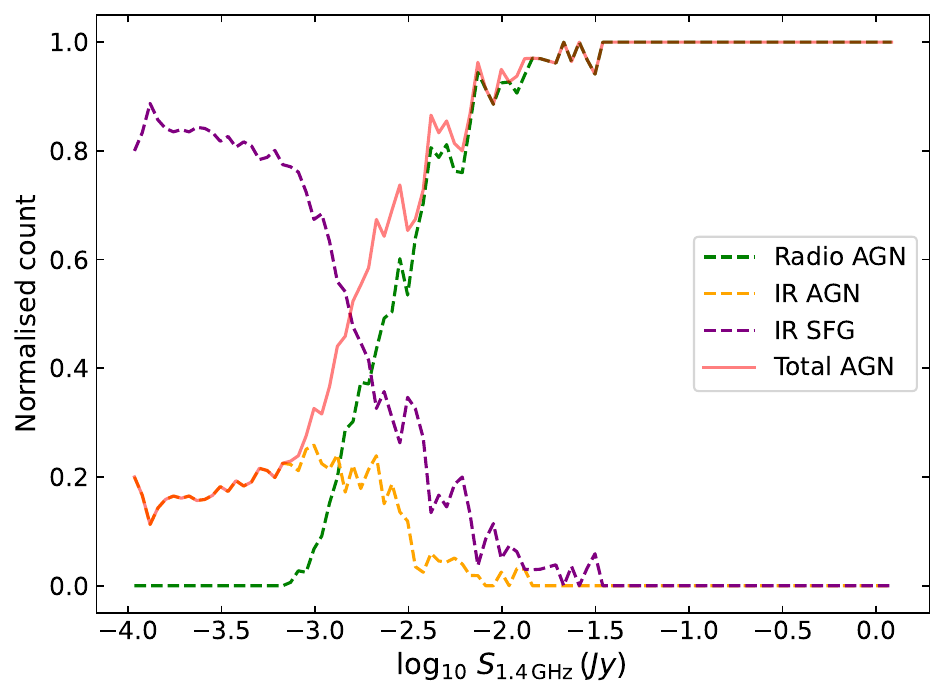}
    \caption{Fractional contribution as a function of radio flux density. The dashed green line represents the radio AGN (green cluster), the dashed orange line represents the IR AGN (orange cluster), the dashed purple line represents the IR SFGs (purple cluster), and the total fraction of AGN at a given radio flux density is represented by the solid red line. It is evident that the star forming fraction is at the maximum below 1\,mJy, above which the radio AGN contributes the most to the radio flux budget.}
    \label{fig:agn_fraction}
\end{figure}

GMM and FCM are found to produce results that can be compared with the astrophysical classifications only for the optical sources. In the case of IR sources, however, no clustering tool other than KMeans seems to work well, and unfortunately, KMeans does not work probabilistically. As an alternative approach, we looked at the normalised number counts of the different populations identified in the IR-radio space shown in Figure\,\ref{fig:ir_radio_clusters}a as a function of \radioflux\,in one hundred bins, shown in Figure\,\ref{fig:agn_fraction}.
\begin{table*}[!ht]
    \centering
    \begin{tabular}{m{4cm}|c|m{8cm}}
    \hline
        Column name & Unit & Description \\
        \hline
        component\_id& - & Radio source identifier taken from the parent G09 and G23 catalogues \\
        uberID&-&GAMA source id\\
        CATAID&-&GAMA source id (from the GAMA II phase)\\
        CatWISE2020\_id&-&CatWISE2020 catalogue id of the radio source\\
        ra\_deg\_cont, dec\_deg\_cont&deg&Right ascension and declination from EMU\\
        ra\_err, dec\_err&arcsec&EMU positional uncertainties\\
        RAcen, Deccen&deg&Right ascension and declination of the optical sources (of the flux-weighted centres, ICRS) from GAMA\\
        RAcen\_err, Deccen\_err&arcsec&GAMA positional uncertainties\\
        RA\_catwise, DEC\_catwise&deg&CatWISE right ascension and declination\\
        RA\_catwise\_err, DEC\_catwise\_err&arcsec&CatWISE2020 positional uncertainties\\
        Total\_flux&milliJansky (mJy)&Integrated radio flux density at 943\,MHz\\
        Total\_flux\_err&milliJansky (mJy)&Uncertainty in the integrated flux density at 943\,MHz\\
        prob\_agn\_gmm&-&Probability for a source to be AGN as per the GMM clustering\\
        prob\_agn\_fcm&-&Probability for a source to be AGN as per the FCM clustering\\
        KMeans\_class&-&KMeans cluster label of the source\\
        GMM\_class&-&GMM cluster label of the source\\
        FCM\_class&-&FCM cluster label of the source\\
        BIRCH\_class&-&BIRCH cluster label of the source\\
        \hline
        \bottomrule
    \end{tabular}
    \caption{The AGN probability (or fraction) catalogue columns and their descriptions. The catalogue contains the GAMA, CatWISE2020, and EMU identifiers, positions, EMU flux densities, GMM and FCM probabilities, and the clustering labels, where 0 corresponds to SFG, and 1 corresponds to AGN.}
    \label{tab:catalogue}
\end{table*}
The fractions are calculated as the ratio of the number of sources in a given class and the total number of sources. The green dashed line corresponds to radio AGN (green cluster), the orange dashed line corresponds to IR AGN (orange cluster), and the purple dashed line corresponds to IR SFG (purple cluster). The solid red line corresponds to the total (IR+radio) AGN fraction. These fractions are different from the probabilities discussed in the case of optical sources. GMM and FCM provide the AGN/SFG fractions in a given source, whereas Figure\,\ref{fig:agn_fraction} gives the AGN/SFG fraction at a given radio flux density for a sample of sources.

It can be seen that the SFGs dominate the sample at lower flux values ($\rm S_{1.4\,GHz}<1\,mJy$), where at the lowest fluxes represent $\approx$\,90\% of the radio source population. This fraction of the SFGs at lower radio flux densities compares well with the existing studies \citep[e.g.][]{2008MNRAS.386.1695S,2017A&A...602A...5N,2021MNRAS.506.3540P,2023MNRAS.523.1729B}. SFGs become the dominant population in the low flux density, most likely because their radio emission is linked to ongoing star formation \citep[e.g.,][]{2008MNRAS.386.1695S}. The dominance of AGN at high radio flux densities can be attributed to extended radio sources and intrinsically high radio luminosities. The remaining $\approx$\,10\% sources at these flux values are IR AGN. Both IR SFGs and AGN start dropping off in numbers above 1\,mJy. An increase in the number of radio AGN is also evident above this flux density. These characteristics are in agreement with the observed trends as discussed above. The IR AGN and SFG (possibly radio-quiet) fractions (orange and purple dashed lines) are still expected to carry some radio emission associated with radio activity. As already stated in \S\,\ref{sec:radio_agn}, even sources without excess radio emission compared to the FIR-radio correlation are found to host radio AGN \citep[see,][]{2019NatAs...3..387P}. We do not provide probabilities of the IR-radio sources similar to \S\,\ref{sec:optical_probs}. Such an analysis would require assessing catalogue completeness and is aimed at future work.

The philosophy of this paper is encapsulated in this statement, `every galaxy is a composite.' When galaxies are classified using a binary scheme, the opportunity to explore the complexity of underlying astrophysical processes is often lost. A probabilistic approach, which quantifies the likelihood of a galaxy being star forming or hosting an AGN, allows for a more nuanced interpretation. This enables deeper astrophysical enquiries, such as investigating transitional populations and mixed-mode activity, all of which are obscured by rigid binary classification. In order to emphasise this perspective, we present a catalogue of G09 and G23 radio sources with optical diagnostic labels and probabilities where possible. The catalogue presents the optical-radio sample with GAMA, CatWISE2020, and EMU identifiers, positions, and the probabilities from GMM and FCM. It would also provide the labels given by the four clustering algorithms, where 0 corresponds to SFG and 1 corresponds to AGN. The details of this catalogue are shown in Table\,\ref{tab:catalogue}. These probabilities can be used for deriving luminosity functions encapsulating the exact astrophysics underneath, and can also be used for training ML models, enabling faster and simpler probabilistic classification (see Carvajal et al. in prep).
\section{Conclusions}\label{sec:conclusion}
Galaxies are commonly considered as a population with binary characteristics when discussing their underlying energy generation mechanisms. This is not the whole picture, as evident from observations and theoretical understandings. The main point of this work is to emphasise that the better question to ask is, ``what percentage of a galaxy's emission is powered by its AGN activity?'' The benefit is that exploring these fractions will lead to a better quantification of astrophysically relevant quantities (e.g. luminosity functions and densities). The number of AGN identified in binary classifications depends strongly on the wavelength regime and the diagnostic used. This is evident from the different numbers of AGN in the optical and IR samples. This means that the more diagnostics we use to explore the population, the larger the AGN sample size will be, urging a multiwavelength analysis in galaxy population studies.

In this work, we evaluated the performance of different unsupervised clustering tools in the optical and IR-radio diagnostic spaces, using an optical-radio sample and an IR-radio sample, respectively. All four clustering algorithms used, namely KMeans, GMM, FCM, and BIRCH, performed well in the optical diagnostic spaces, whereas only KMeans worked well for the IR-radio diagnostic spaces. We found that the classifications and clusters agree with each other for up to 90\% of the time in the case of SFGs and up to 80\% of the time in the case of AGN, in both the optical and IR-radio diagnostics. Radio flux density as an additional parameter in the optical diagnostic space did not seem to impact the results. This is probably caused by the sample selection, where the requirement of BPT and other optical lines discards the absorption line galaxies. The IR-radio diagnostics revealed quite interesting results, where one out of the three clusters in the IR space occupied the radio AGN region in the radio diagnostic space.

We used this feature to propose two novel IR-radio AGN diagnostics, $(i)$ a two-dimensional space using the $\rm W_1-W_2$ colour and \radioflux. This diagnostic selects radio AGN with $\approx$\,90\% completeness and $\approx$\,90\% reliability, $(ii)$ a three-dimensional space that uses the $\rm W_2$ magnitude as a third dimension to the two-dimensional diagnostic, which also selects radio AGN with $\approx$\,90\% reliability and $\approx$\,90\% completeness, combining the best of both the spectral regimes. The superiority of the three-dimensional radio AGN selection is a crucial outcome of this work. Apart from being driven by the mixed parameter spaces of IR and radio, and thereby capturing the essence of both the wavelength regimes, the advent of wider and deeper radio surveys like EMU necessitates the development of new and improved selection criteria for galaxies active in the radio regime. We also publish a catalogue of the G09 and G23 radio sources with the associated probabilities of them being an AGN in the optical regime, emphasising the acute need for considering galaxies to be composed of probabilistic contributions rather than a binary population.

\begin{acknowledgement}
This scientific work uses data obtained from Inyarrimanha Ilgari Bundara / the Murchison Radio-astronomy Observatory. We acknowledge the Wajarri Yamaji People as the Traditional Owners and native title holders of the Observatory site. The Australian SKA Pathfinder is part of the Australia Telescope National Facility (\url{https://ror.org/05qajvd42}), which is managed by CSIRO. Operation of ASKAP is funded by the Australian Government with support from the National Collaborative Research Infrastructure Strategy. ASKAP uses the resources of the Pawsey Supercomputing Centre. Establishment of ASKAP, the Murchison Radio-astronomy Observatory and the Pawsey Supercomputing Centre are initiatives of the Australian Government, with support from the Government of Western Australia and the Science and Industry Endowment Fund. This paper includes archived data obtained through the CSIRO ASKAP Science Data Archive, CASDA (\url{http://data.csiro.au}).

GAMA is a joint European-Australasian project based around a spectroscopic campaign using the Anglo-Australian Telescope. The GAMA input catalogue is based on data taken from the Sloan Digital Sky Survey and the UKIRT Infrared Deep Sky Survey. Complementary imaging of the GAMA regions is being obtained by a number of independent survey programmes including GALEX MIS, VST KiDS, VISTA VIKING, WISE, Herschel-ATLAS, GMRT and ASKAP, providing UV to radio coverage. GAMA is funded by the STFC (UK), the ARC (Australia), the AAO, and the participating institutions. The GAMA website is \url{http://www.gama-survey.org/}.
\end{acknowledgement}

% PASA uses footnotes, not endnotes. \endnote in this template will behave like \footnote; and \printendnotes will not output anything.
% \printendnotes

\bibliography{bibtemplate}

%\appendix

\end{document}